\documentclass[twocolumn,times]{aastex61}
\usepackage{times}
\usepackage{epsfig}
\usepackage{amsmath, amsthm, amssymb}
\usepackage{color}
\usepackage{microtype}
\usepackage{float}
\usepackage{verbatim}
\usepackage{wrapfig}
\hypersetup{backref,breaklinks,colorlinks,citecolor=blue}
\usepackage[all]{hypcap}
\usepackage{xspace}

\newcommand{\beq}{\begin{equation}}
\newcommand{\eeq}{\end{equation}}
\newcommand{\bea}{\begin{eqnarray}}
\newcommand{\eea}{\end{eqnarray}}

\newcommand{\msun}{$M_\sun$\xspace}

\newcommand{\FLASH}{{\tt FLASH}}
\newcommand{\mesa}{{\tt mesa20}}
\newcommand{\mesalr}{{\tt mesa20\_LR}}
\newcommand{\mesapert}{{\tt mesa20\_pert}}
\newcommand{\mesalrpert}{{\tt mesa20\_LR\_pert}}
\newcommand{\mesaoct}{{\tt mesa20\_oct}}
\newcommand{\mesalroct}{{\tt mesa20\_LR\_oct}}
\newcommand{\mesavlr}{{\tt mesa20\_v\_LR}}
\newcommand{\mesavlroct}{{\tt mesa20\_v\_LR\_oct}}
\newcommand{\mesatwod}{{\tt mesa20\_2D}}
\newcommand{\mesatwodpert}{{\tt mesa20\_2D\_pert}}
\newcommand{\mesaoned}{{\tt mesa20\_1D}}

\def\fref{Fig.~\ref}
\def\eref{Eq.~\ref}
\def\sref{\S~\ref}

\shorttitle{3D CCSNe}
\shortauthors{O'Connor \& Couch}

\begin{document}

\title{Exploring Fundamentally Three-dimensional Phenomena in High-fidelity Simulations of Core-collapse Supernovae}

\author[0000-0002-8228-796X]{Evan P. O'Connor}
\affiliation{Department of Astronomy and the Oskar Klein Centre, Stockholm
  University, AlbaNova, SE-106 91 Stockholm, Sweden \href{mailto:evan.oconnor@astro.su.se}{evan.oconnor@astro.su.se}}

\author[0000-0002-5080-5996]{Sean M. Couch}
\affiliation{Department of Physics and Astronomy, Michigan State University, East Lansing, MI 48824, USA; \href{mailto:couch@pa.msu.edu}{couch@pa.msu.edu}}
\affil{Department of Computational Mathematics, Science, and Engineering, Michigan State University, East Lansing, MI 48824}
\affil{National Superconducting Cyclotron Laboratory, Michigan State University, East Lansing, MI 48824, USA}
\affil{Joint Institute for Nuclear Astrophysics-Center for the Evolution of the Elements, Michigan State University, East Lansing, MI 48824, USA}

\begin{abstract}

  The details of the physical mechanism that drives core-collapse
  supernovae (CCSNe) remain uncertain.  While there is an emerging
  consensus on the qualitative outcome of detailed CCSN mechanism
  simulations in 2D, only recently have high-fidelity 3D simulations
  become possible.  Here we present the results of an extensive set of
  3D CCSN simulations using high-fidelity multidimensional neutrino
  transport, high-resolution hydrodynamics, and approximate general
  relativistic gravity.  We employ a state-of-the-art 20 $M_\sun$
  progenitor generated using the Modules for Experiments in Stellar
  Astrophysics \citep[MESA;][]{farmer:2016, paxton:2011, paxton:2013,
    paxton:2015, paxton:2018} and the SFHo equation of state of
  \citet{Steiner:2013}.  While none of our 3D CCSN simulations explode
  within $\sim$500\,ms after core bounce, we find that the presence of
  large scale aspherical motion in the Si and O shells surrounding the
  collapsing iron core aid shock expansion and bring the models closer
  to the threshold of explosion.  We also find some dependence on
  resolution and geometry (octant vs. full $4\pi$). As has been noted
  in other recent works, we find that the post-shock turbulence plays
  an important role in determining the overall dynamical evolution of
  our simulations.  We find a strong standing accretion shock
  instability (SASI) that develops at late times during the shock
  recession epoch.  The SASI aids in transient shock expansion phases,
  but is not enough to result in shock revival.  We also report that
  for a subset of our simulations, we find conclusive evidence for the
  Lepton-number Emission Self-sustained Asymmetry (LESA) first
  reported in \cite{Tamborra:2014}, but until now, not confirmed by
  other simulation codes. Both the progenitor asphericities and the
  SASI-induced transient shock expansion phases generate transient
  gravitational waves and neutrino signal modulations via
  perturbations of the protoneutron star by turbulent motions.


\end{abstract}

\section{Introduction}
\label{sec:intro}

At the ends of their lives, stars more massive than about 8--10 \msun
develop inert, neutrino-cooled iron cores in their centers.  The iron
core in such stars builds up over the period of days due to core
silicon burning and, ultimately, silicon shell burning until it
reaches the effective Chandrasekhar mass, which in general depends on
the electron fraction, $Y_e$, and entropy of the core
\citep{Woosley:2002}.  Once the iron core reaches this critical mass,
gravitational collapse ensues.  The collapse of the core is
accelerated by electron captures onto nuclei and protons, the rates of
both of which increase as the collapse drives the densities, and
therefore the electron chemical potential higher and higher.  The
collapse accelerates to an appreciable fraction of the speed of light.
The inner 0.5 \msun of the core is in sonic contact and collapses
homologously while the remaining outer core collapse supersonically.
Electron-type neutrinos produced by the rapid electron captures are
eventually ``trapped'' in the inner core once densities exceed about
$10^{12}$\,g\,cm$^{-3}$. When the collapsing core reaches nuclear
densities the residual strong force, also known as the nuclear force,
between the nucleons starts to dramatically resist the force of
gravity.  This halts the collapse of the inner core which then
elastically rebounds and launches a pressure wave that quickly
steepens into the supernova shock.  The supernova shock propagates out
into the still-infalling outer core leaving in its wake a hot
protoneutron star (PNS).  Energy losses from neutrino emission in the
hot layers of the PNS and from the dissociation of the iron-group
nuclei at the shock cause the shock to initially stall and become an
accretion shock.  At this time, the beginnings of neutrino-driven
convection in the neutrino-heated, convectively-unstable layers behind
the supernova shock break the spherical symmetry that has otherwise
dominated the evolution so far. These multidimensional instabilities
are thought to be the crucial piece of the puzzle that assists the
canonical explosion mechanism, the neutrino mechanism
\citep{bethe:1985}, to reenergize the supernova shock and drive the
supernova explosion.

Tremendous progress in the theoretical study of the CCSN mechanism has
been made in recent years.  This progress has been spurred largely by
the advent of high-fidelity fully-3D simulations that can both capture
the crucial hydrodynamic instabilities that form soon after core
bounce and model the detailed neutrino transport that drives the
thermodynamic evolution of the PNS and the shock-heating layers.
\citep{Hanke:2013, melson:2015, melson:2015a, lentz:2015,
  Roberts:2016, ott:2018, summa:2018}.  Three dimensional simulations
also allow for the ability to directly simulate precollapse progenitor
asphericities.  At the point of collapse and outside of the iron core,
there are shells of lighter elements. Immediately outside the iron
core there is typically silicon shell that is undergoing nuclear
burning that may or may not be convective at the point of core collapse, depending on the preceding
evolution \citep[cf.,][]{chieffi:2013, sukhbold:2014}, above this silicon
shell is an oxygen shell that is also typically convective. Convection
in these shells imprints non-spherical density structure and velocity
motions on the progenitor.  These are thought to be crucial for not
only correctly simulating the core-collapse evolution, but perhaps
needed in order to obtain explosions themselves. 
Evidence for this was seen as early as the 2D work of \citet{burrows:1996}, but \citet{couch:2013b} were the first to show definitively on the basis of 3D simulations that non-spherical structure in the shells surrounding the collapsing iron core could qualitatively alter the outcome of CCSN simulations. 
\cite{Muller:2015} also showed, using parameterized
precollapse perturbations in 2D simulations, that progenitor asphericities can play a
strong role in the development of turbulence and lateral kinetic
energy in the gain region, which boosts the effectiveness of neutrino
heating, supports a larger shock radius, and helps drive
explosions. 
Going beyond parameterized, artificially-imposed perturbations, \citet{couch:2015} simulated the final few minutes of evolution through iron core collapse in a 15 \msun star, directly calculating several convective turn over times in the Si-burning and O-burning shells. 
Resulting CCSN mechanism simulations with this 3D progenitor model showed a modest increase in the strength of the explosion, though not the strong qualitative impact found by \citet{couch:2013b}.
Subsequently, \cite{muller:2016a} simulated a longer period of evolution through core collapse in an 18 \msun star, finding substantial large-scale non-spherical motion in the O-burning shell.
CCSN mechanism simulations with this progenitor \citep{muller:2017} showed a dramatic impact from the realistic 3D progenitor, yielding a successful explosion for a case in which the comparable 1D progenitor failed. 

In this work, we contribute new \FLASH{}, high-resolution, full 3D,
energy-dependent, three-species neutrino-radiation-hydrodynamic
simulations of a 20\,$M_\odot$ zero-age main sequence (ZAMS) mass
progenitor star evolved using the Modules for Experiments in Stellar
Astrophysics \citep[MESA;][]{farmer:2016, paxton:2011, paxton:2013,
  paxton:2015, paxton:2018} software. In our eight 3D simulations, we
explore not only the 3D evolution of this progenitor, but we also
explore the impact of progenitor asphericities, resolution,
octant/full 3D symmetry, dimensionality, and variations in the
evolution due to the neutrino transport physics. We do not obtain
explosions in any of our 3D simulations, however we are able to
quantitatively comment on the impact of each of the above aspects on
the explosion mechanism. Our precollapse progenitor asphericities,
particularly those in the silicon shell, are very effective at driving
the development of turbulence in the postshock layers. While only
transitory in nature, this leads to larger average shock radii, more
neutrino heating, and stronger turbulence.  In the majority of our
full 3D simulations we see the presence of the standing accretion
shock instability (SASI). While this arises only once the supernova
explosions is seemingly failing, we note that the growth of the
standing accretion shock instability at late times can lead to epochs
of shock reenergization. Unfortunately, these excursions are never
large enough to drive an explosion in the time we have simulated. We
see an imprint of the SASI motion on the neutrino luminosities. In one
of our full 3D simulations we find a strong presence of the
lepton-number emission self-sustained asymmetry (LESA). The LESA was
first seen in \cite{Tamborra:2014}, but until now, has yet to be
confirmed by other simulation codes. Lastly, we report on the
gravitational wave signatures from all of our full 3D simulations. We
find that the accretion of progenitor asphericities excite the PNS and
lead to a temporary growth of the gravitational wave strength, as does
the collapse of SASI spiral waves.  The presence of such features in
an observation of a nearby CCSNe may be a unique signal to the
presence of strong transient turbulence in the accretion flow.

This paper is organized as follows.  In the next section,
\sref{sec:methods}, we introduce the \FLASH{} code and describe the
hydrodynamics, gravity, and neutrino transport.  We also discuss our
initial conditions and the parameterization we use for our precollapse
progenitor asphericities. In \sref{sec:results}, we first present an
overview of our eight 3D simulations, contrasting them against each
other.  We explore in detail the nature of the presence of the SASI
and the LESA.  We also present the neutrino and gravitational wave
signals from our simulations.  We discuss and conclude in
\sref{sec:conclusions}.

\section{Methods}
\label{sec:methods}

\subsection{Hydrodynamics and Gravity}

We make use of the \FLASH{} hydrodynamics framework
\citep{fryxell:2000,dubey:2009} that we have outfitted for CCSNe in
\cite{couch:2013c,couch:2013a,couch:2014} and more recently have
included both multidimensional, energy-dependent neutrino-radiation
transport based on the moment formalism and effective general
relativistic gravity \cite{oconnor:2018}. For these simulations we use
\FLASH{}'s unsplit hydrodynamic solver that makes use of the piecewise
parabolic method (PPM; \citealt{colella:1984}) and the hybrid HLLC
Riemann solver, which reduces to HLLE in the presence of
shocks. During each \FLASH{} timestep, we solve for the new
hydrodynamic state (the $(n+1)$ state) before the neutrino transport
step. We refer the reader to previous works for a summary of the
hydrodynamics and gravity methods
\citep{couch:2013c,couch:2013a,couch:2014,oconnor:2018} and instead
focus on a presentation of the multidimensional neutrino transport.

\subsection{Neutrino Transport}

Neutrinos play a critical role in core-collapse supernovae.  First and
foremost, they provide a tremendous cooling channel for the newly formed
protoneutron star, they also are a source of heat in the shocked
layers above the protoneutron star, the so-called gain region. This
heat source is thought to be critical for the development of the explosion.
Following the evolution of the neutrinos from the core of the
protoneutron star out through the gain region required a complex
treatment of neutrino radiation transport. Furthermore, since the opacity of
the matter has a strong dependence on the neutrino energy, this
transport must be done in an energy-dependent way.  In \FLASH{}, we
have implemented a multidimensional, multispecies, energy-dependent
two-moment (with an analytic closure) neutrino-radiation transport
scheme. It is based on the work of \cite{oconnor:2015, shibata:2011,
  cardall:2013}. We have outlined our \FLASH{} implementation in great
detail in \cite{oconnor:2018}.  For the majority of our 3D simulations
presented here, we ignore the velocity dependence of the neutrino
transport equations (i.e. we set $\vec{v}$, and its derivatives to
0). We also ignore the gravitational redshift term that moves
neutrinos between energy bins, although we keep the overall source
term from the gravitational redshift.  These approximations are due to
not having fully implemented this physics prior to beginning our 3D
simulations.  For comparison purposes, we do perform two simulations
with full velocity and gravitational red-shift dependence which were
started after these improvement were made to our neutrino transport
code. For completeness, we show the form of our moment evolution
equations here and refer the reader to the Appendix of
\cite{oconnor:2018} for implementation details. The coordinate frame,
energy-dependent, neutrino energy density ($E$; zeroth moment)
evolution equation in Cartesian coordinates is given as
\begin{eqnarray}
\nonumber
&&\partial_t E + \partial_i [\alpha F^i] - \partial_\nu \left[\alpha \nu (L^{ij}\partial_iv_j +
    F^i\partial_i\phi)\right]\\
 &&\hspace*{0.2cm}=\alpha (W [\eta - \kappa_a J] -
    [\kappa_a+\kappa_s]H^t -  F^i\partial_i \phi)\,,
\end{eqnarray}
and the coordinate frame, energy-dependent, neutrino momentum density
($F^i$; first moment) evolution equation is
\begin{eqnarray}
 \nonumber && \partial_t F^i + \partial_j [\alpha P^{ij}]
- \partial_\nu \left[\alpha \nu (N^{ijk}\partial_jv_k +
  P^{ij}\partial_j\phi)\right]\\
&&\hspace*{0.2cm}= -\alpha([\kappa_s +\kappa_a]H^i - W [\eta -\kappa_a
J]v^i + E \partial_i \phi) \,,
\label{eq:Fi}
\end{eqnarray}
\noindent
where in these equations, $\vec{v}$ is the matter velocity (with the
associated Lorentz factor of $W$); $\alpha = \exp{(\phi)}$ is the lapse
(with $\phi$ being the gravitational potential), $\nu$ and
$\partial_\nu$ denote the neutrino energy and the energy-space
derivative; $\kappa_a$, $\kappa_s$, and $\eta$ are the matter
absorption opacity, scattering opacity, and emissivity, respectively;
$P^{ij}$ is the second moment of the neutrino distribution function,
that we use an analytic closure for; $L^{ij}$ and $N^{ijk}$ are higher
moment tensors used in the energy-space flux determination; and $J$
and $H^i$ are the fluid-frame zeroth and first moments that can be
expressed in terms of $E$, $F^i$, and $P^{ij}$,
\begin{eqnarray}
J &= &W^2[ E - 2 F^i v_i + v^i v^j P_{ij}]\,,\\
H^t &=& W^3[ - ( E- F^i v_i) v^2 + F^i v_i - v_i v_j P^{ij}]\,,\\
H^i &= &W^3[ -( E -  F^j v_j) v^i + F^i -  v_j P^{ij}]\,.
\end{eqnarray}
\noindent
These equations are energy and species dependent.  We use 12 energy
groups and three neutrino species ($\nu_e$, $\bar{\nu}_e$, and
$\nu_x = \{\nu_\mu + \bar{\nu}_\mu + \nu_\tau +
\bar{\nu}_\tau\}$). This amounts to 144 evolution equation. Since we
perform the spatial flux and energy flux calculations explicitly,
these differential equations are only coupled within an energy group
and spatial zone.  We perform the resulting $4\times4$ matrix
inversions analytically.

The neutrino-matter interaction coefficients, $\kappa_a$, $\kappa_s$,
and $\eta$ are also energy and species dependent. They also depend on the
matter density, temperature, and electron fraction.  We use {\tt NuLib}
\citep{oconnor:2015} to generate a three dimensional table (in
density, temperature, and electron fraction) which we then
tri-linearly interpolate, for each energy group and species, on the fly using the
$(n+1)$-state matter variables from the hydrodynamic step. For the
underlying neutrino interactions, we use isotropic scattering on
nucleons, alpha particles, and heavy nuclei following
\cite{bruenn:1985,burrows:2006a} with corrections for weak magnetism
and recoil from \cite{horowitz:2002}. For emission and absorption processes
involving electron type neutrinos and antineutrinos, we employ charged
current interactions on neutron, protons, and heavy nuclei,
specifically,
\begin{eqnarray}
n + e^+ & \leftrightarrow & p + \bar{\nu}_e\,, \\
p + e^- & \leftrightarrow & n + \nu_e\,,\\
(A,Z) + e^- & \leftrightarrow & (A,Z-1) + \nu_e\,,
\end{eqnarray}
\noindent
from \cite{bruenn:1985} using weak magnetism and recoil corrections
from \cite{horowitz:2002}.  Finally, for thermal pair-processes we use
both electron-positron annihilation and nucleon-nucleon bremsstrahlung
following \cite{burrows:2006a} as implemented in \cite{oconnor:2015}.
We only include these processes for heavy-lepton neutrinos.

\subsection{Initial Conditions}
\label{sec:IC}

For our suite of 3D simulations, we adopt the $20\,M_\odot$, solar
metallicity progenitor model from \cite{farmer:2016} produced using a non-equilibrium 204 isotope nuclear network with Modules for
Experiments in Stellar Astrophysics \citep[MESA;][]{paxton:2011,
  paxton:2013, paxton:2015, paxton:2018}.  We use the SFHo nuclear
equation of state (EOS) \citep{Steiner:2013,hempel:2012,oconnor:2010} everywhere.
We simulate the collapse phase and early (first 15\,ms) postbounce
phase using GR1D \citep{oconnor:2015} and then transition to FLASH.
We use the same equation of state table and neutrino interactions as
well as the same description of gravity (i.e. using a general
relativistic effective potential instead of full GR) which ensures a
smooth transition between codes.  We do interpolate from the 18 energy
groups used in GR1D to the 12 energy groups used in the 3D FLASH
simulations.

It is worth discussing the MESA 20\,$M_\odot$ progenitor and comparing
it to another model frequently used in the literature.  The
$20\,M_\odot$ MESA model (referred to as MESA20) is fairly similar to
the 20\,$M_\odot$ model from \cite{Woosley:2007} (referred to here as
s20). The compactness ($\xi_{M}$; \citealt{oconnor:2011}) of the two
20\,$M_\odot$ models are also similar:
$\xi_{1.75}^\mathrm{MESA20} \sim 0.69$ and
$\xi_{1.75}^\mathrm{s20} \sim 0.75$. The silicon-oxygen interface,
where a sharp drop in density occurs ($\sim$50\% is both models), is
located at a radius of $\sim$2400\,km and $\sim$2600\,km for the
MESA20 and s20, respectively.  The baryonic mass coordinates of these
interfaces are $\sim$1.75\,$M_\odot$ and $\sim$1.8\,$M_\odot$,
respectively.

We simulate the postbounce phase in 3D using a Cartesian grid with
adaptive mesh refinement. Our smallest grid zones are
$\sim$488\,m. The adaptive mesh refinement would ensure that all of
the postshock region is fully refined, however, for our baseline
simulations we limit the refinement so as to only maintain a minimum
effective angular resolution of $\Delta\,x/r \lesssim 0.009$ or
$\Delta\,x/r \lesssim 0^\circ.53$.  This restriction leads to a
reduction in resolution from $\Delta x \sim 488\,$m to
$\Delta x \sim 1\,$km at $r \sim 108\,$km, and further reduction by a
factor of 2 at $r \sim 216\,$km. We refer to this as our standard
resolution.  We also have lower resolution simulations (denoted with a
{\tt LR} in the model name) where we still take the smallest
$\Delta x = 488\,$m, but only enforce $\Delta\,x/r \lesssim 0.015$ or
$\Delta\,x/r \lesssim 0^\circ.88$.  This leads to the first resolution
decrement at $r \sim 65$\,km, and subsequent decrements at
$r\sim 130\,$km, $r\sim 260\,$km, etc. Our baseline simulations are
full 3D (4$\pi$), however we perform some simulations in octant symmetry
(denoted with a {\tt oct} in the model name). For two of our
simulations we use an improved neutrino transport that includes full
velocity dependence.  We denote these simulations with a {\tt v} in
the model name.  Finally, in two of our simulations we introduce
velocity perturbations into the silicon and oxygen shell at the time
of mapping to FLASH (denoted with a {\tt pert} in the model name).  We
describe the details of these perturbations below. For the simulations
without progenitor perturbations we do not impose any seed
perturbations to the simulations, instead, we let the Cartesian grid
seed deviations from spherical symmetry.

In total, we perform eight 3D simulations.  Our flagship simulations,
\mesa{} and \mesapert{}, use our standard resolution.  We also perform
these simulations at lower resolution: \mesalr{} and \mesalrpert{}. We
include velocity dependence in \mesavlr{}. Finally, we do three octant
simulations \mesaoct{}, \mesalroct{}, and \mesavlroct{}. We also
perform a collection of 1D and 2D simulations in order to address the
dimensional dependence.

\subsubsection{Progenitor Perturbations}

For two of our simulations we impose three-dimensional precollapse
velocity asphericities onto the matter at the beginning of the
simulations.  We base the strength of the perturbations off of the
convective velocities in the original 20$M_\odot$ progenitor model
\cite{farmer:2016}. Our model has two regions of interest that are
convective at the point of core-collapse.  The silicon shell and the
oxygen shell.  Since we do the collapse in 1D and map to our 3D code
at 15\,ms after bounce, we must wait to apply the perturbations. We
then assign the perturbations based on the mass coordinate of the
convective zones in the progenitor model. We use the methods of
\cite{Muller:2015} to determine the perturbations to the velocity
field.  We include only perturbations in $v_r$ and $v_\theta$ and keep
$\delta v_\phi=0$, though we do include a sinusoidal $\phi$ dependence in
the perturbed velocities. At the time of mapping, the silicon shell is
located between $r_\mathrm{min} = 1.25\times10^8$\,cm and
$r_\mathrm{max} = 1.99\times10^8$\,cm.  In the notation of
\cite{Muller:2015}, for the silicon shell we take $n=1$, $\ell=9$, and
$m=5$.  For the overall scaling factor, we take
$C=10.5\times10^{30}$g\,s$^{-1}$. For the convective oxygen shell we
take $r_\mathrm{min} = 2.23\times10^8$\,cm and
$r_\mathrm{max} = 6.5\times10^8$\,cm, $n=1$, $\ell=5$, and $m=3$, and
an overall scaling factor of $C=8.4\times10^{30}$g\,s$^{-1}$. To
extend the equations of \cite{Muller:2015} to 3D, we keep the
$\phi$ dependence in the spherical harmonic functions of their Eq. 7
and opt to use the real part of their Eq. 8.  These factors give maximum
Mach numbers of 0.3 and 0.2 for the lateral motions in both the
silicon and oxygen shells, respectively. The perturbations in the
radial velocities have a similar magnitude (i.e.
$|(\delta v)_r| \sim 0.2 c_s$, but are placed on the background
velocity field, which has a infalling Mach number of $\sim$0.8 in the
silicon shell and $\sim$0.4 at the base of the oxygen shell.

\section{Results}
\label{sec:results}
\subsection{Overview}

Our suite of eight 3D simulations; which all use the same progenitor
model, nuclear equation of state, and neutrino opacities; span a
number of interesting parameters including resolution, geometry,
inclusion of progenitor perturbations, and inclusion of
neutrino-transport velocity dependence.  We have done this in order to
be able to test the sensitivity of our 3D simulations to each of these
aspects.  In this section we will present an overview of all our of 3D
results and discuss each of these conditions in turn. 

\begin{figure*}[tb]
  \plotone{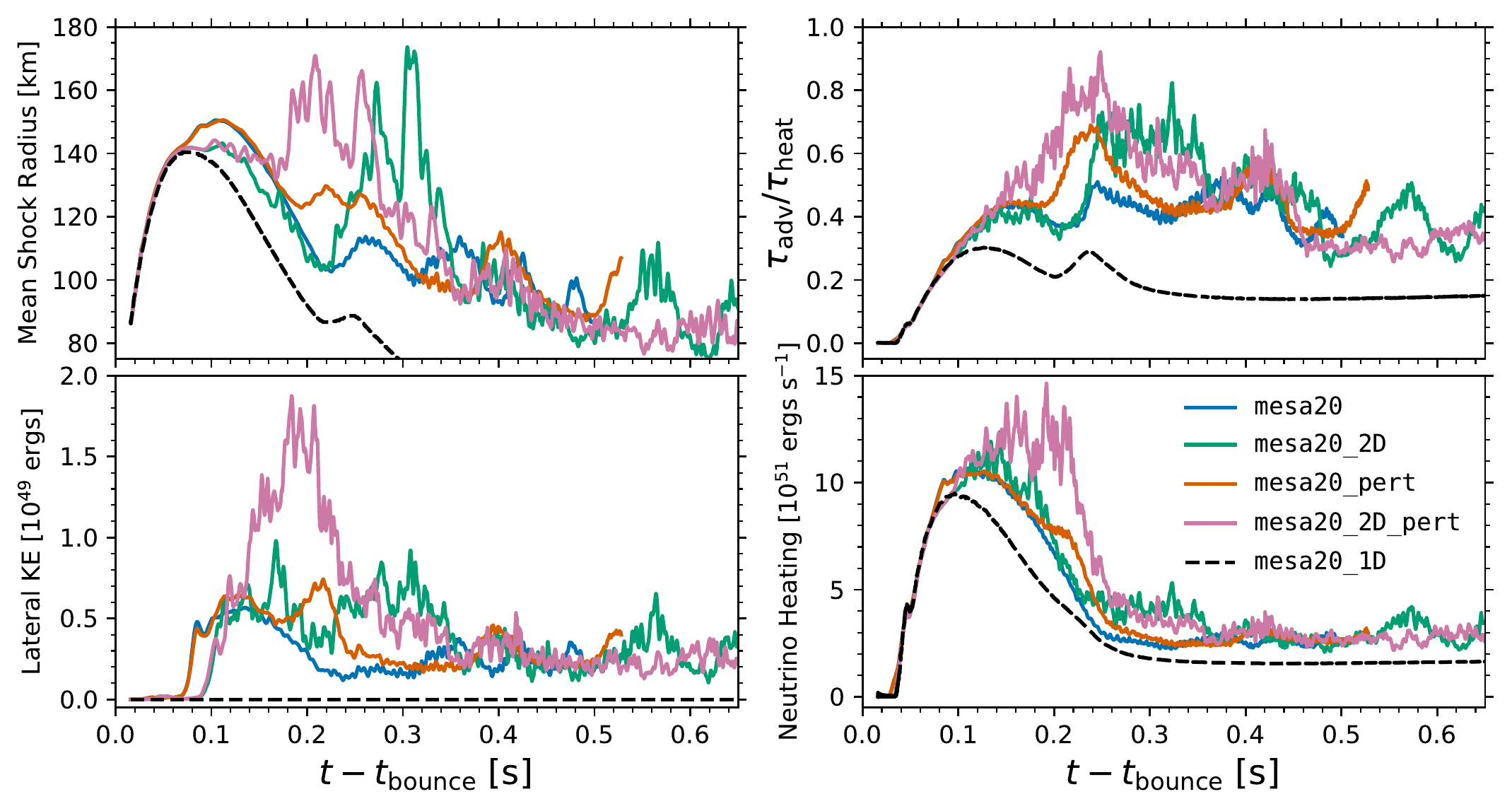}
  \caption{$r_\mathrm{sh}$ (top left), $T^\mathrm{lat}_\mathrm{gain}$
    (bottom left), $\tau_\mathrm{adv}/\tau_\mathrm{heat}$ (top right),
    and $\dot{Q}_\mathrm{heat}$ (bottom right) vs. post-bounce time
    for models \mesa{} (blue), \mesatwod{} (green), \mesapert{} (red),
    \mesatwodpert{} (pink), and \mesaoned{} (black).}\label{fig:mesa20vsdim}
\end{figure*}

We will discuss each of 3D simulations below with the help of
\fref{fig:mesa20vspert} through \fref{fig:mesa20vsdim}. In each of
these figures we will show four basic quantities which we describe
here. \emph{1)} The mean shock radius $r_\mathrm{sh}$; \emph{2)} the
lateral kinetic energy in the gain region
$T^\mathrm{lat}_\mathrm{gain}$, which is the sum of all non-radial
kinetic energy in the gain region; \emph{3)} the ratio of the
advection timescale through the gain region to the heating timescale
in the gain region,
\begin{equation}
\frac{\tau_\mathrm{adv}}{\tau_\mathrm{heat}} =
\frac{M_\mathrm{gain}/\dot{M}}{|E_\mathrm{gain}|/\dot{Q}_\mathrm{heat}}\,,\label{eq:tadvtheat}
\end{equation}
where $M_\mathrm{gain}$ is the mass of the gain region, $\dot{M}$ is
the accretion rate outside the shock, $E_\mathrm{gain}$ is the energy
of the matter in the gain region (gravitational + kinetic + internal
[relative to free neutrons]), and $\dot{Q}_\mathrm{heat}$ is the rate
of energy injection via neutrino heating; and \emph{4)} the heating
rate in the gain region, $\dot{Q}_\mathrm{heat}$. The ratio,
$\tau_\mathrm{adv}/\tau_\mathrm{heat}$, gives a quantitative measure
of how close a simulation is to an explosion, especially when directly
comparing models with the same underlying progenitor model and
equation of state.  Empirically it has been found that when this ratio
reaches 1 an explosion is expected
\citep{marek:2009, oconnor:2018}. Physically,
$\tau_\mathrm{adv}/\tau_\mathrm{heat}=1$ means that the time it takes
to change the energy of the matter in the gain region by of order
itself via neutrino heating is equal to the time it takes to accrete through the gain
region. In this time the matter can be heated and unbound before
settling onto the protoneutron star.

\subsubsection{Impact of an Aspherical Progenitor}
\label{sec:perturbations}

\begin{figure*}[tb]
  \plotone{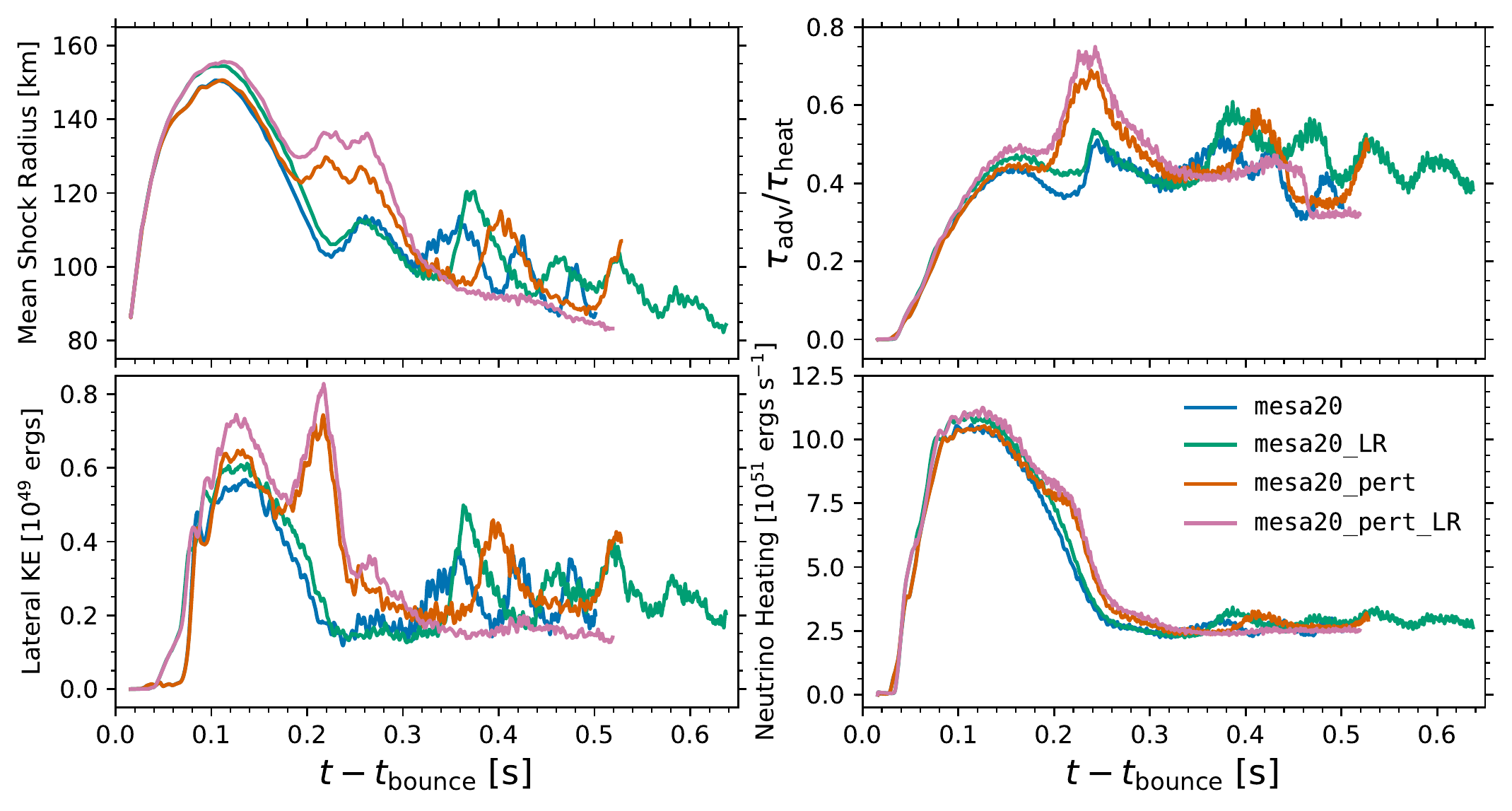}
  \caption{$r_\mathrm{sh}$ (top left), $T^\mathrm{lat}_\mathrm{gain}$
    (bottom left), $\tau_\mathrm{adv}/\tau_\mathrm{heat}$ (top right),
    and $\dot{Q}_\mathrm{heat}$ (bottom right) vs. post-bounce time
    for models \mesa{} (blue), \mesapert{} (red), and their low
    resolution variants \mesalr{} (green) and \mesalrpert{}
    (pink). The striking difference is the impact of the progenitor
    perturbations (from the silicon shell) at $\sim$200\,ms after
    bounce.  These perturbations give an increased shock radius,
    increased lateral kinetic energy, and increased neutrino heating
    around this time.  The \mesapert{} and \mesalrpert{} simulations
    are quantitatively closer to explosion.}\label{fig:mesa20vspert}
\end{figure*}

\begin{figure*}[tb]
  \includegraphics[width=0.24\textwidth]{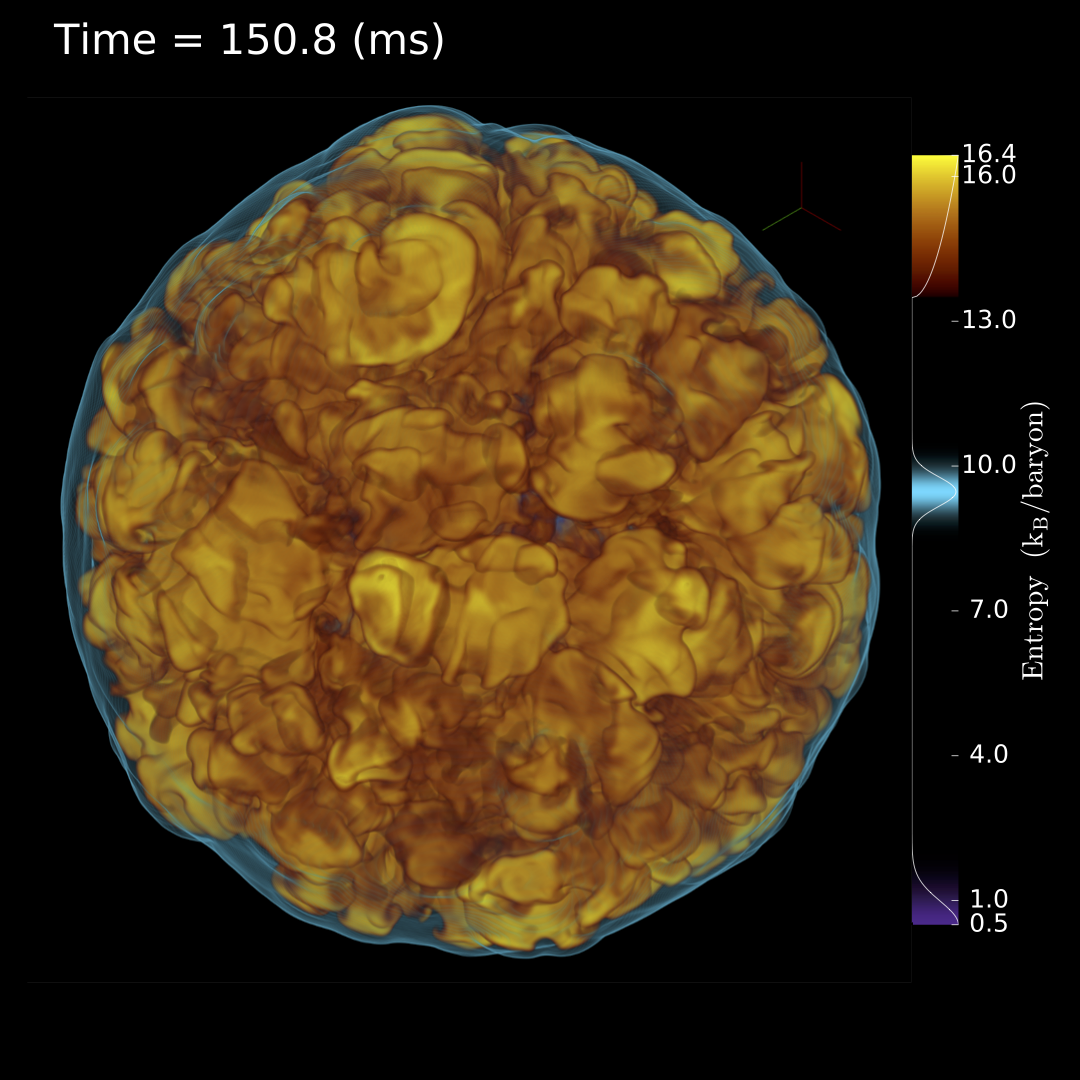}
  \includegraphics[width=0.24\textwidth]{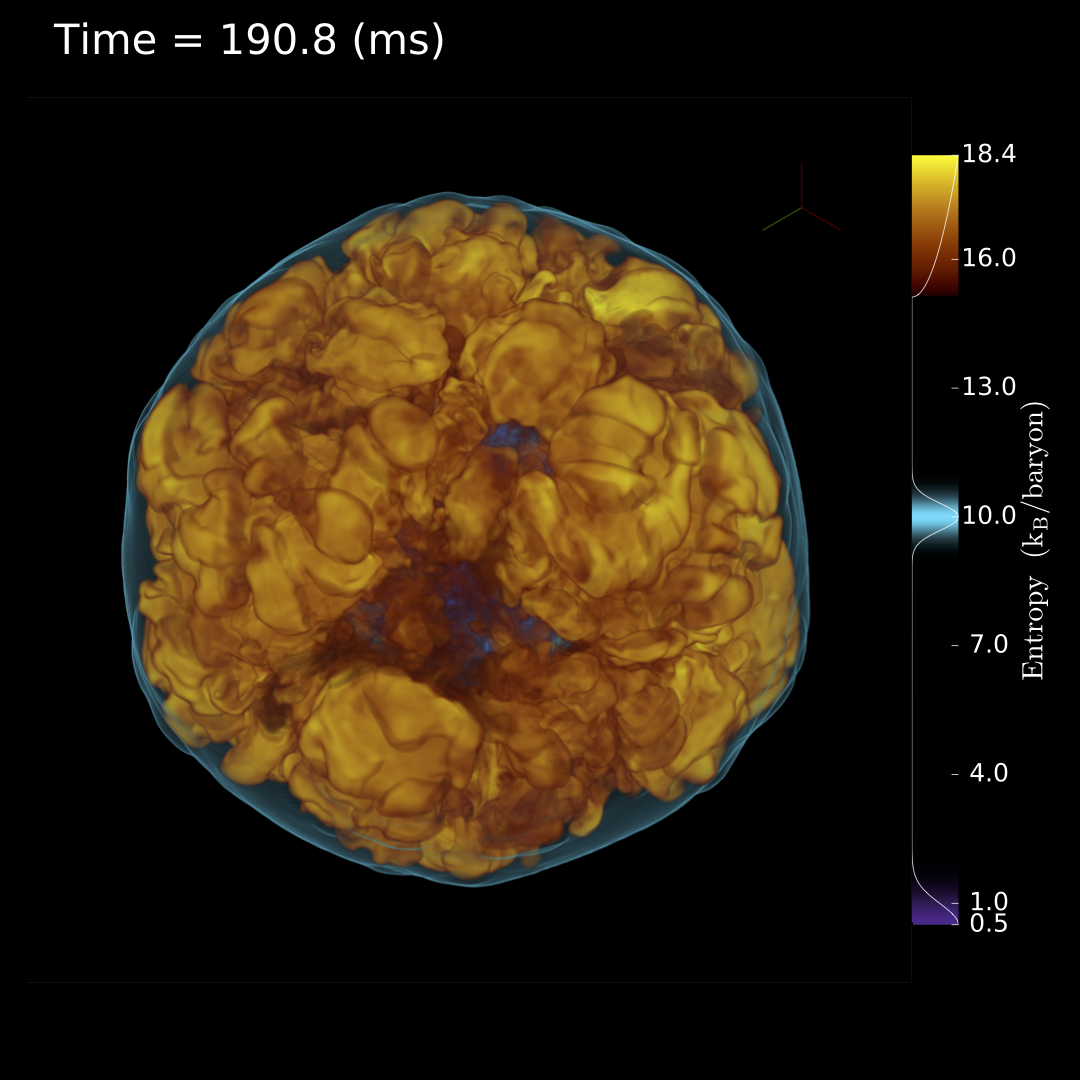}
  \includegraphics[width=0.24\textwidth]{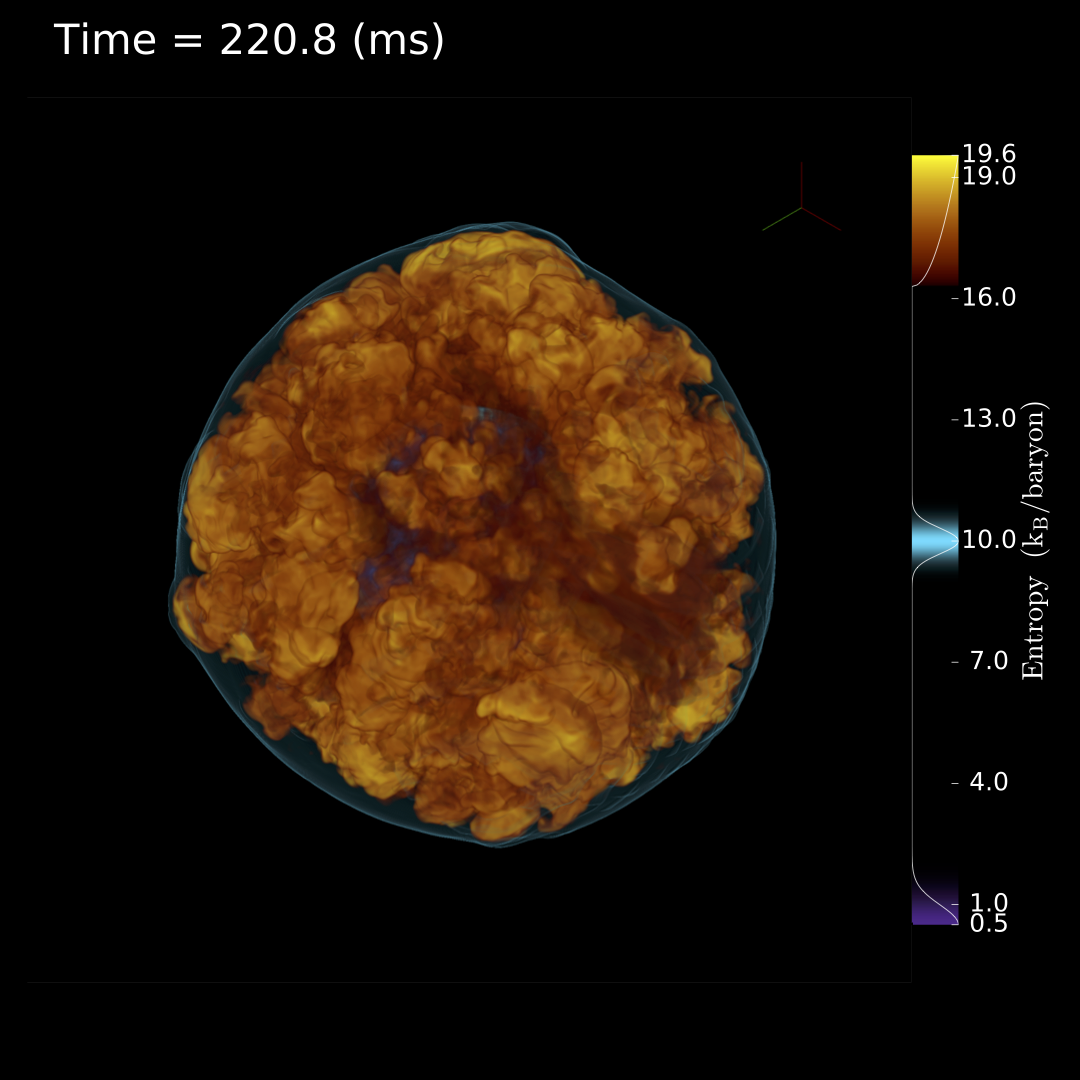}
  \includegraphics[width=0.24\textwidth]{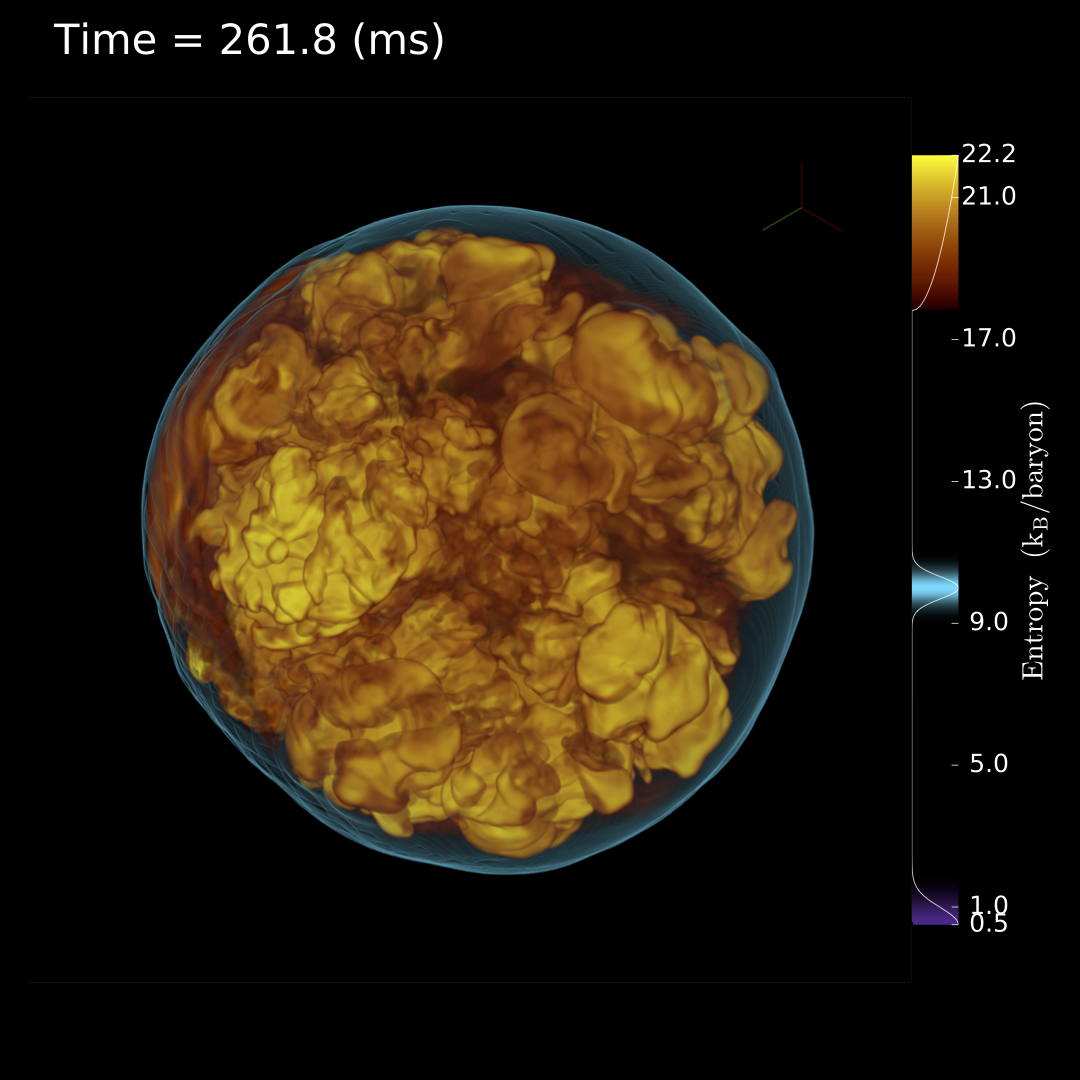}\\
  \includegraphics[width=0.24\textwidth]{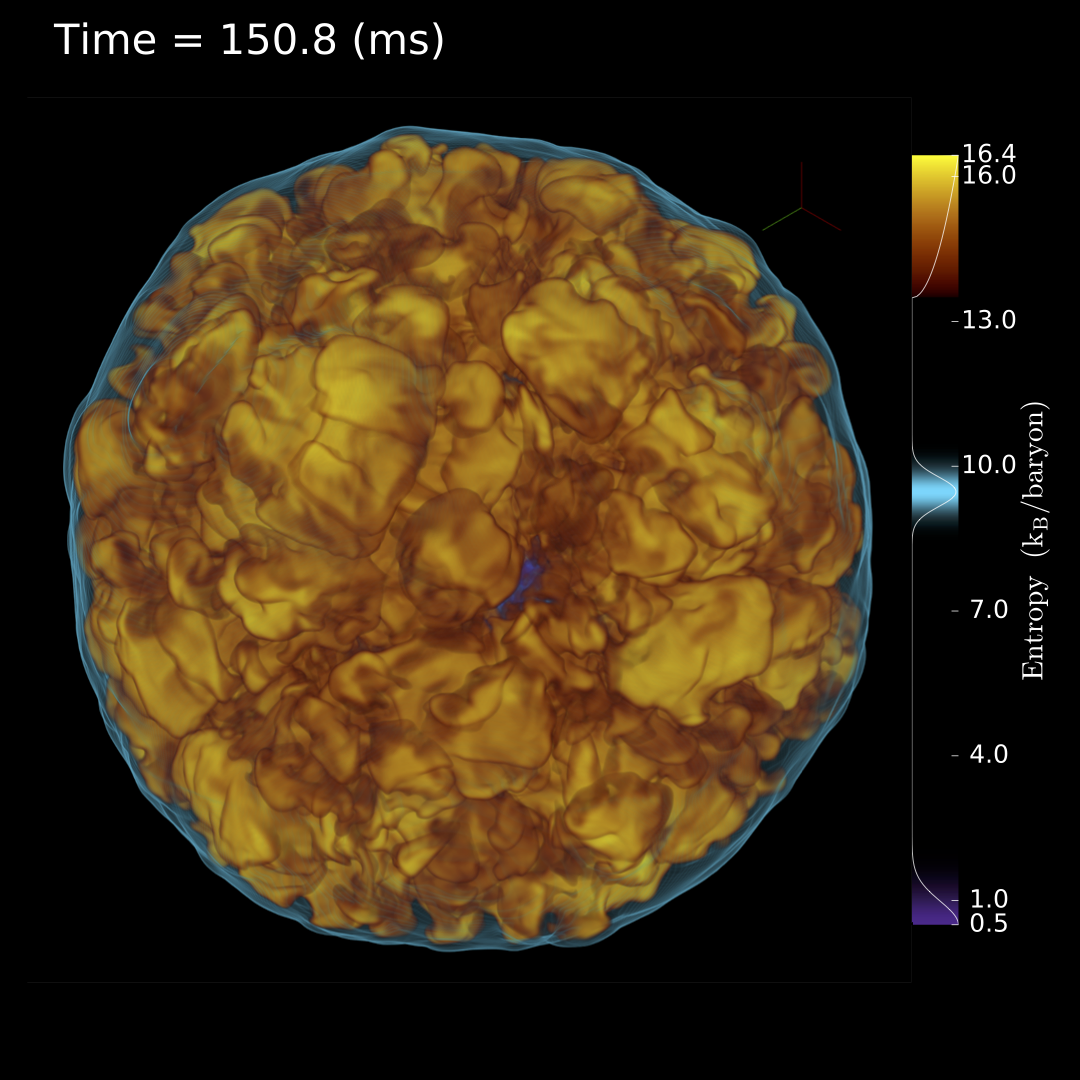}
  \includegraphics[width=0.24\textwidth]{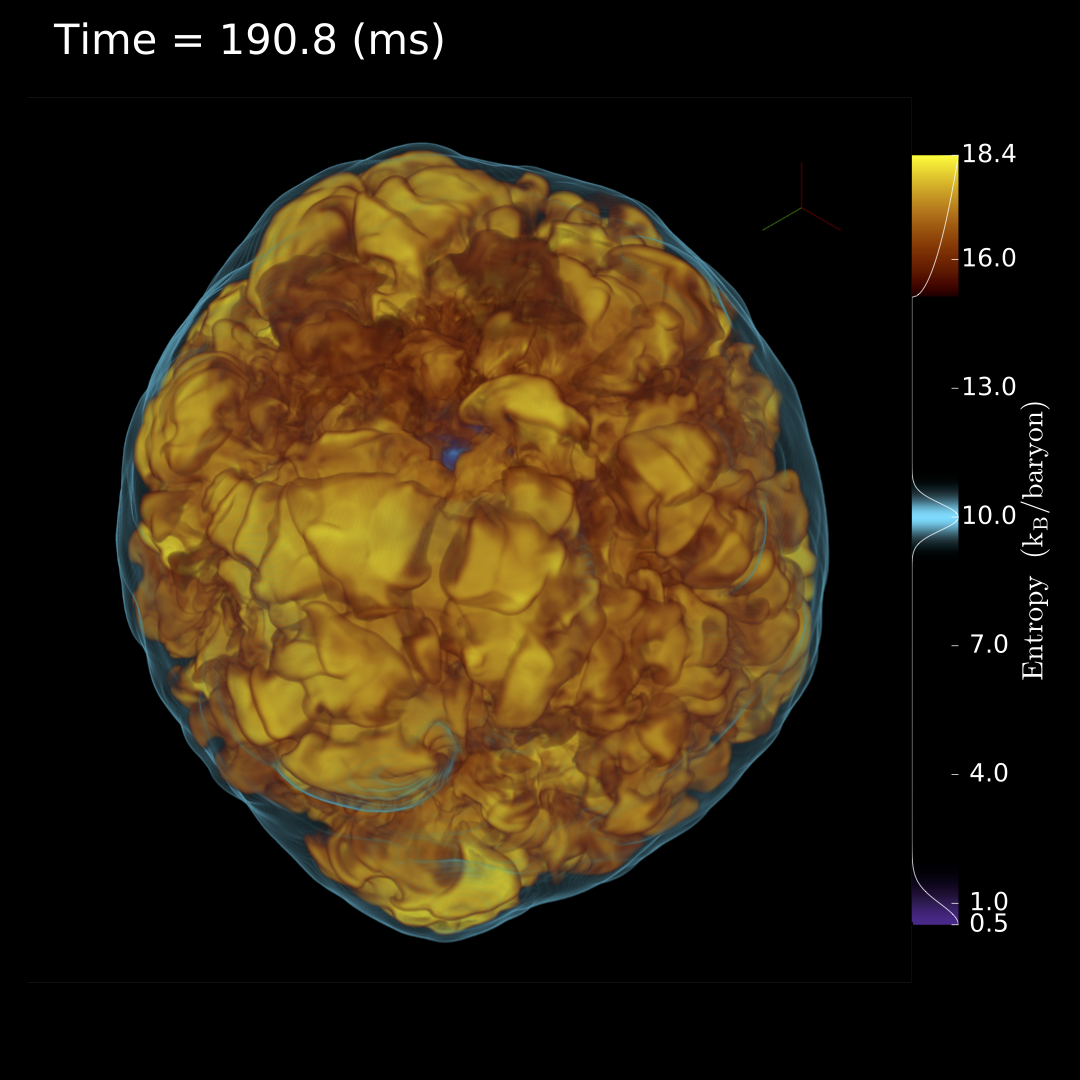}
  \includegraphics[width=0.24\textwidth]{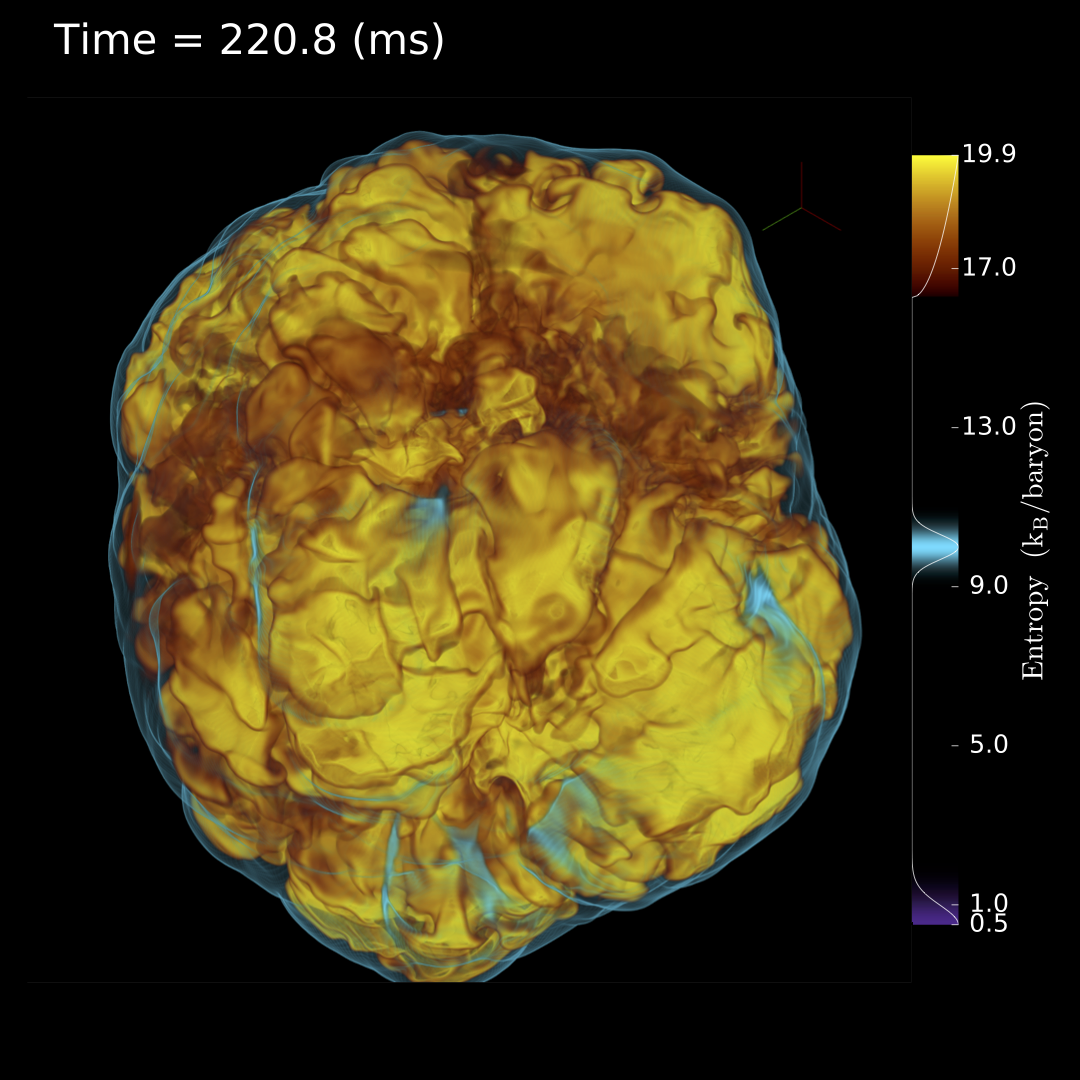}
  \includegraphics[width=0.24\textwidth]{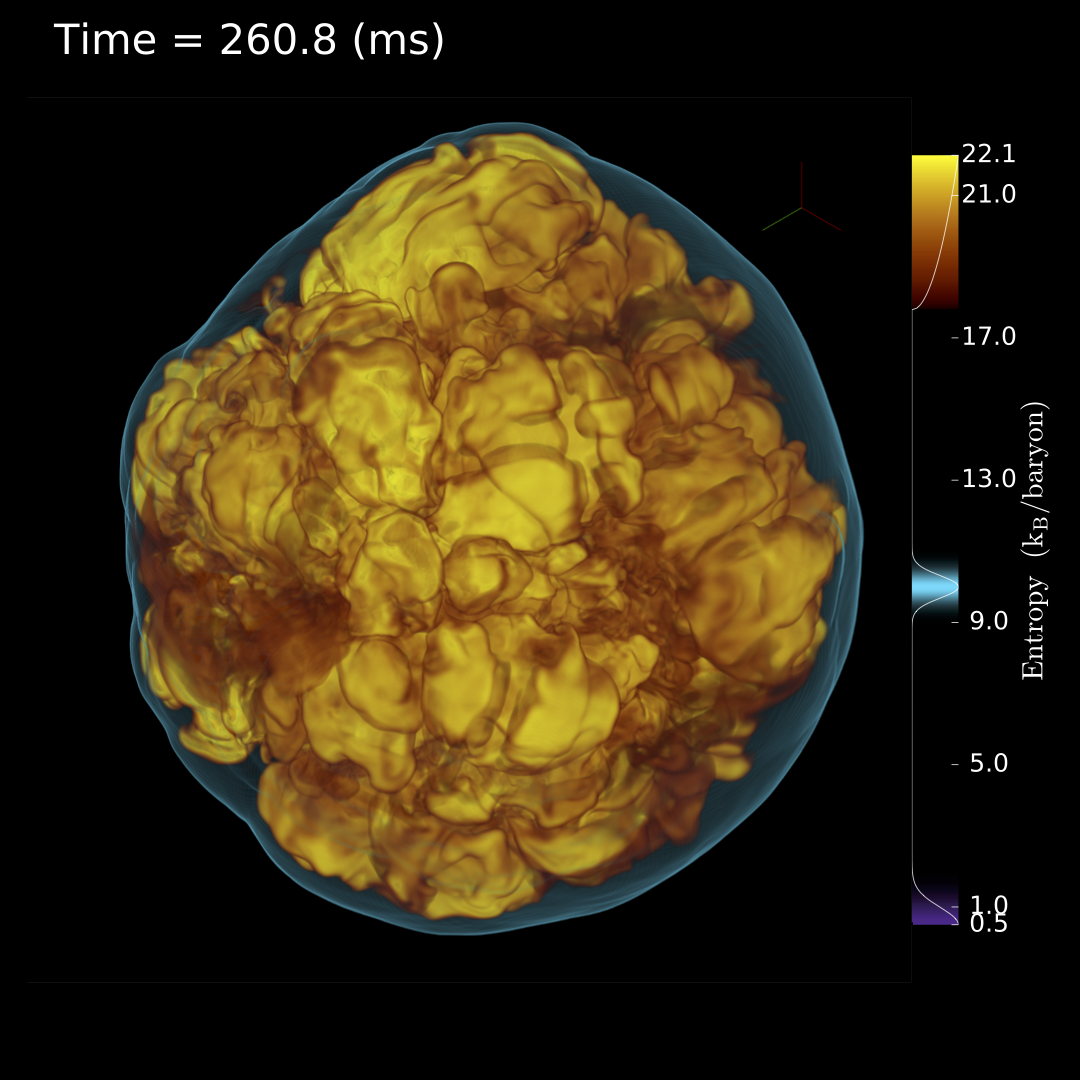}
  \caption{Volume renderings of \mesa{} and \mesapert{} with focus on
    the entropy variations in the gain region for four postbounce
    times of $\sim$150\,ms, $\sim$190\,ms, $\sim$220\,ms, and
    $\sim$260\,ms.  The supernova shock is denoted by the thin cyan
    surface, and the PNS is denoted by the opaque magenta sphere in
    the center (not always visible). These renderings show the impact
    of the imposed progenitor perturbations that accrete through the shock
    starting around 180\,ms and have a maximal effect around
    220\,ms. By 260\,ms, these perturbations have accreted out of the
    gain region and the simulations become more similar.
  }\label{fig:vrs}
\end{figure*}

To begin, we discuss our two flagship simulations: \mesa{} and
\mesapert{}. These differ in the inclusion of progenitor perturbations
in the silicon and oxygen shell as presented in \sref{sec:IC}. In
\fref{fig:mesa20vspert}, we show the four basic quantities described
above, $r_\mathrm{sh}$ (top left), $T^\mathrm{lat}_\mathrm{gain}$
(bottom left), $\tau_\mathrm{adv}/\tau_\mathrm{heat}$ (top right), and
$\dot{Q}_\mathrm{heat}$ (bottom right).  Also in this figure we show
the low resolution variants of each of these simulations, \mesalr{}
and \mesalrpert{}, which will be primarily discussed in the following
section.  We show volume renderings at $\sim$150\,ms, $\sim$190\,ms,
$\sim$220\,ms, and $\sim$260\,ms (from left the right) in
\fref{fig:vrs} for both the \mesa{} (top) and
\mesapert{} (bottom) simulations to aid our discussion. Immediately we can see the impact of the aspherical
progenitor in \mesapert{} and \mesalrpert{}.  Starting as early as
$\sim$100\,ms after bounce we see the first signs of the imposed
asphericity in the silicon shell impacting the postbounce
dynamics. The amount of lateral kinetic energy in the aspherical
models is diverging from and significantly exceeding the spherical
models. The impact on the shock radius and the neutrino heating is not
appreciable at this time.  The mass accretion rate is high at this
time, the turbulent motions that do form are quickly accreted out of
the gain region.  It is not until later, $\sim$180\,ms, that we see an
impact on these other measures.  Since this time is well before when
the interface between the silicon and oxygen shells accretes through
the shock (at $\sim$250\,ms), the behavior seen at this time is still
due to the perturbations imposed on the silicon shell. At this time,
we see a dramatic rise in the lateral kinetic energy
($\sim7\times 10^{48}$\,erg\,s$^{-1}$ for \mesapert{} compared to
$\sim2\times 10^{48}$\,erg\,s$^{-1}$ for \mesa{} at 220\,ms), the
shock recession temporarily ceases, and the neutrino heating rate is
enhanced, $7.5\times 10^{51}$\,erg\,s$^{-1}$ in \mesapert{} compared
to $5\times 10^{51}$\,erg\,s$^{-1}$ in \mesa{} at $\sim$220\,ms. We
note that the increase in the lateral kinetic energy is not because
there is significantly more lateral kinetic energy accreting through
the shock in the aspherical models (there is not), but rather that the
non-radial velocities entering the post-shock region are much more
effective at seeding the convective instabilities that drive turbulence.

The quantitative impact of the progenitor asphericity is clear in the
graph of the $\tau_\mathrm{adv}/\tau_\mathrm{heat}$. During the phase
where the silicon shell perturbations are accreting through the gain
region both the gain region itself is larger (giving a long advection
time) and the neutrino heating rate is higher (giving a smaller
heating timescale).  We note that the longer advection time dominates
the contribution to the larger
$\tau_\mathrm{adv}/\tau_\mathrm{heat}$. After the silicon-oxygen shell
interface accretes through the shock the perturbations from the
silicon shell cease and are quickly accreted through the gain region
and onto the PNS.  We find that the perturbations imposed on the
oxygen shell do not have the same qualitative impact as those from the
silicon shell.  We see no significant excess of lateral kinetic energy
during this phase and the evolution of the \mesapert{} simulation
begins to qualitatively match that of \mesa{}. The shock radius
quickly recedes, and the lateral kinetic energy and the neutrino
heating drop.  $\tau_\mathrm{adv}/\tau_\mathrm{heat}$ also drops from
its peak value of $\sim$0.7 to $\sim$0.45.

\subsubsection{Impact of Reduced Resolution}

From \fref{fig:mesa20vspert}, we can also assess the impact of
resolution. For our flagship simulations, \mesa{} and \mesapert{}, we
also perform lower resolution simulations with and without progenitor
perturbations: \mesalr{} and \mesalrpert{}. These lower resolution
simulations have the same central zone size ($\sim$488\,m) but we
enforce $\Delta\,x/r \lesssim 0.015$ (or
$\Delta\,x/r \lesssim 0^\circ.88$) outside of 60\,km instead of
$\Delta\,x/r \lesssim 0.009$ (or $\Delta\,x/r \lesssim 0^\circ.53$)
outside of 90\,km. The lower resolution grid seeds strong numerical
perturbations when the shock passes our fixed refinement levels. This
gives rise to earlier turbulence (and increased lateral kinetic
energy) in the low resolution simulation compared to the standard
resolution.  As a result, the lower resolution simulations also give
slightly larger shock radii, $\sim$5\,km or 3\%, when they eventually
stall at $\sim$100\,ms after bounce and begin to recede.  Not until
the standard resolution simulations experience strong SASI motions at
$\sim$300\,ms and $\sim$350\,ms for \mesa{} and \mesapert{},
respectively, does the shock overtake that of the lower resolution
simulations. This is discussed more in \sref{sec:SASI}.

The early time differences between the two resolutions follows closely
the results from \cite{Roberts:2016}, who also perform 3D
neutrino-radiation transport simulations with two different
resolutions. In the lower resolution simulations here, and in
\cite{Roberts:2016}, turbulence develops earlier and is most likely
responsible for the slightly larger initial shock radius.  After the
turbulence becomes fully developed, both the simulations here and in
\cite{Roberts:2016} see very little difference in the turbulent nature
of the two simulations with different resolutions.  The various
components of the Reynolds stress show no systematic differences with
resolution, except at the earliest times. Both the low and high
resolution,  full 3D, simulations in \cite{Roberts:2016} are
successful, however the low resolution one always maintains a larger
radius and explodes earlier than the high resolution simulation. While
none of our simulations explode, the lower resolution simulations are
quantitatively closer to explosion as seen in
$\tau_\mathrm{adv}/\tau_\mathrm{heat}$ from
\fref{fig:mesa20vspert}. We note that our low resolution simulation
($\Delta\,x \sim 2$\,km between $\sim$130\,km and $\sim$260\,km) lies
between the high and low resolution simulations of \cite{Roberts:2016}
and our high resolution simulation is higher ($\Delta\,x \sim 1$\,km
between $\sim$110\,km and $\sim$220\,km) than that of
\cite{Roberts:2016}.

\subsubsection{Impact of Octant Geometry}

\begin{figure*}[tb]
  \plotone{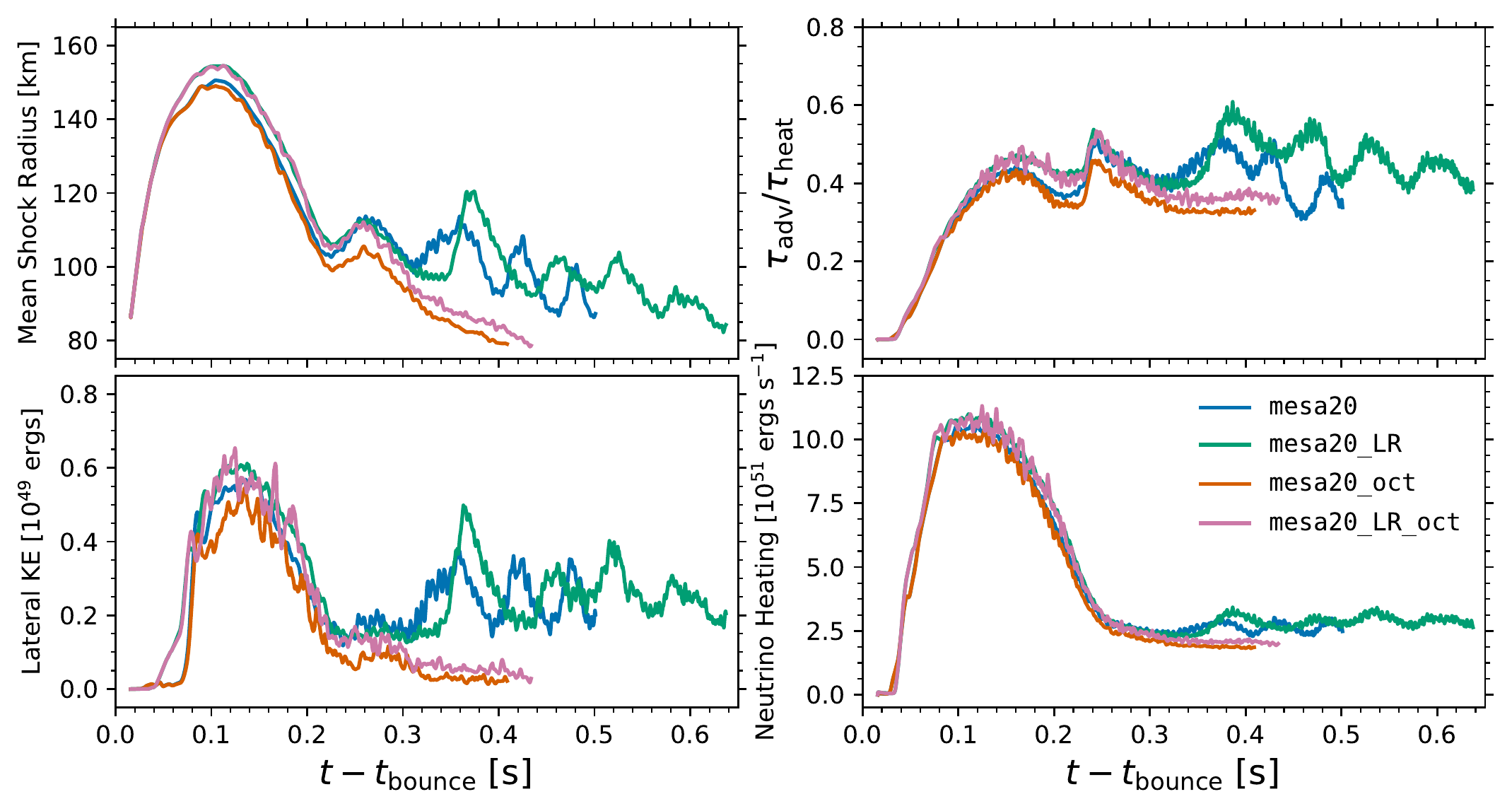}
  \caption{$r_\mathrm{sh}$ (top left), $T^\mathrm{lat}_\mathrm{gain}$
    (bottom left), $\tau_\mathrm{adv}/\tau_\mathrm{heat}$ (top right),
    and $\dot{Q}_\mathrm{heat}$ (bottom right) vs. post-bounce time
    for models \mesa{} (blue), \mesalr{} (green), and their octant
    variants \mesaoct{} (red) and \mesalroct{} (pink). Note, the
    lateral kinetic energy and the neutrino heating from \mesaoct{}
    and \mesalroct{} have each been multiplied by 8 in order to
    account for the octant symmetry.}\label{fig:mesa20vsoct}
\end{figure*}

In \fref{fig:mesa20vsoct}, we show the effect of imposing octant
symmetry on our simulations. We show our main simulation, \mesa{}
(blue), as well as its low resolution variant \mesalr{} (green).  For
each of these, we show variants, \mesaoct{} (red) and \mesalroct{}
(pink), where we restrict the motion to the first octant by imposing
reflecting boundary conditions on each of the main axis. In the early
evolution, prior to $\sim$100\,ms, we see very little impact of the
octant symmetry. Until $\sim$200\,ms, difference that do arise are
smaller than the difference observed between the standard and low
resolution simulations.  For \mesaoct{}, we see a small decrease in
the lateral kinetic energy and well as the neutrino heating, shock
radius, and $\tau_\mathrm{adv}/\tau_\mathrm{heat}$. This is similar to
the effect seen in \cite{Roberts:2016} who also perform octant
simulations. Since the octant symmetry prevents the growth of the
largest scale modes that help support the shock, after $\sim$100\,ms
both the octant simulations of \cite{Roberts:2016} and the octant
simulations presented here have marginally smaller average shock radii
and neutrino heating.  For both the standard and low resolution
simulations we note that the increased noise for the quantities from
the octant simulations in \fref{fig:mesa20vsoct} is due to the smaller
number of zones that we average over (by a factor of 8).

After 200\,ms, the difference between the full 3D simulations and the
octant variants becomes more interesting.  Due to the octant symmetry,
global modes are suppressed. As a result, our octant simulations
behave much like failed 1D models.  While the full 3D simulations
experience repeated episodes of SASI growth accompanied by excursions
of the mean shock radius, the shock radii in the octant models are
rapidly receding. The heating is also notably reduced, at times up to
$\sim$25\% lower, from the full 3D models, consistent with the smaller
shock radii. From the ratio of $\tau_\mathrm{adv}/\tau_\mathrm{heat}$,
the octant simulations are quantitatively the farthest from explosion.

\subsubsection{Impact of Velocity-dependent Transport}
\label{sec:vdep} 

\begin{figure*}[tb]
  \plotone{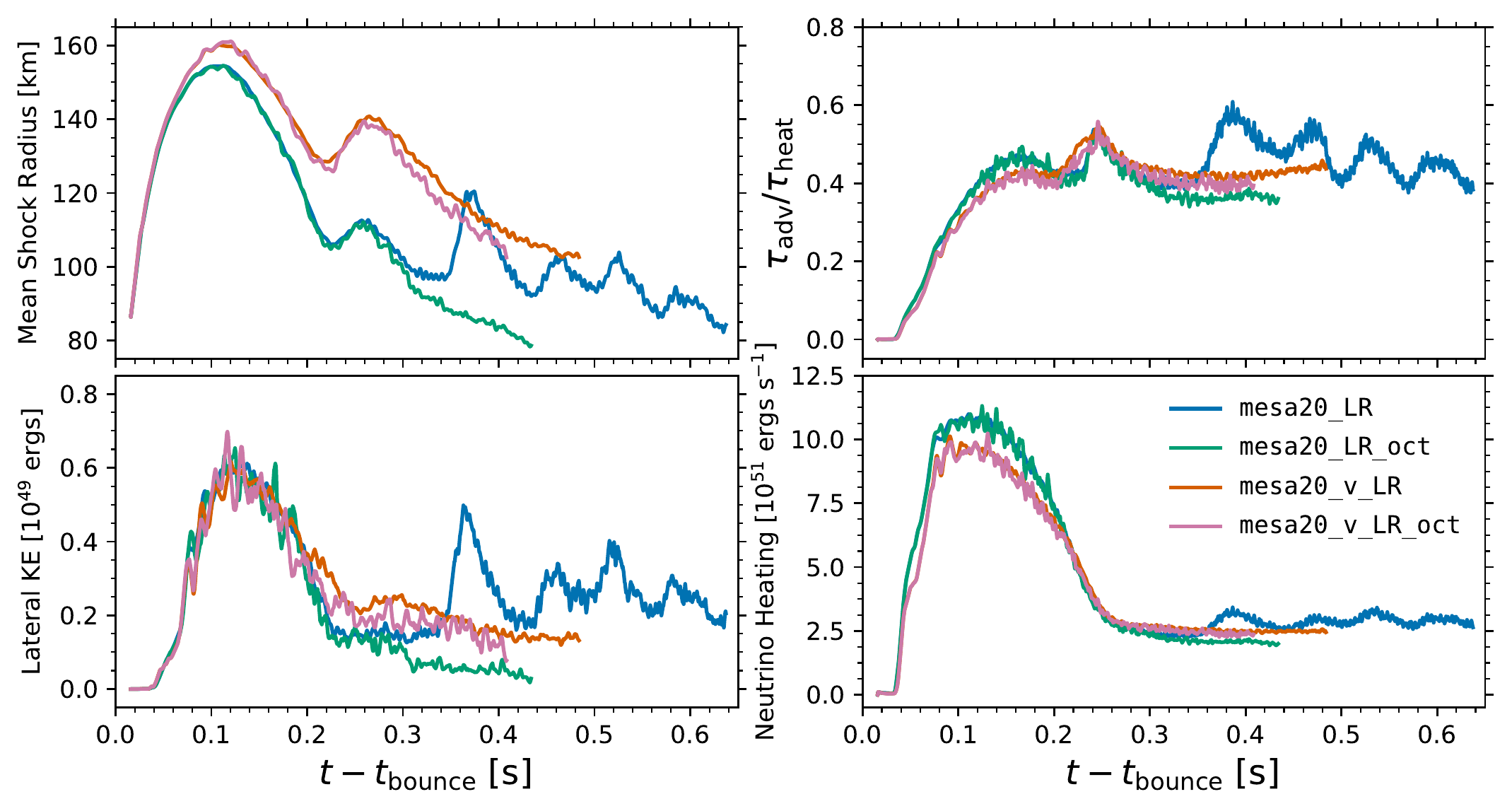}
  \caption{$r_\mathrm{sh}$ (top left), $T^\mathrm{lat}_\mathrm{gain}$
    (bottom left), $\tau_\mathrm{adv}/\tau_\mathrm{heat}$ (top right),
    and $\dot{Q}_\mathrm{heat}$ (bottom right) vs. post-bounce time
    for models \mesalr{} (blue), \mesalroct{} (green), and their octant
    variants \mesavlr{} (red) and \mesavlroct{} (pink). Note, the
    lateral kinetic energy and the neutrino heating from \mesalroct{}
    and \mesavlroct{} have each been multiplied by 8 in order to
    account for the octant symmetry.}\label{fig:vdep}
\end{figure*}

We also perform simulations using a velocity-dependent variant of our
neutrino-transport. Unlike the other simulations presented here, these
simulations fully account for velocity dependent effects in the
transport, gravitational redshift of the neutrinos as they stream out
of the gravitational well, as well as advection of neutrinos with the
fluid flow when they are trapped. We perform two low resolution
simulations with velocity dependence, one in full 3D, \mesavlr{} and
one in octant symmetry \mesavlroct{}.  We show the results of these
simulations in \fref{fig:vdep}, where we also include the non-velocity
dependent simulations \mesalr{} and \mesalroct{}. We find that
including velocity dependence has several effects on the evolution.  

In the velocity dependent results, there appears to be slightly less
overall heating ($\sim$10\%) in the gain region before $\sim$200\,ms
after bounce.  This is due to, unfortunately, slightly different
definitions of the energy exchange term in our two simulations that
cannot be corrected with post-processing, but is purely a book-keeping
difference.  In the velocity-dependent simulations, our matter
exchange term is recorded as the rate of total energy exchange with
the matter. While in the non-velocity-dependent simulations we record
the rate of internal energy exchange with the matter. The difference
between these two rates is the energy change due to momentum
absorption of the neutrinos, which in the case of the gain region,
acts to reduce the kinetic energy of the matter (hence why this rate
is lower).  We note that in both cases we properly take into account
the energy exchange due to momentum absorption for the hydrodynamic
source term. In 1D simulations we observe that the heating rate
derived from the rate of total energy exchange in both the
velocity-dependent and non-velocity-dependent simulations are very
similar. This similarity is naively not what one would expect, but is
a coincidental cancelation of two effects that seemingly equally
impact the amount of neutrino heating with opposite signs.  The first
effect is from the gravitational redshift, which reduces the energy of
the neutrinos, hence the effectiveness of neutrino capture in the gain
region.  A competing effect is that the background flow of the
material in the gain region is against the neutrino flow so that in
the frame of the fluid, the neutrinos are boosted to higher energies.

One of the more visible impacts of the added velocity dependence seen
in \fref{fig:vdep} is the evolution of the shock radius.  With
velocity dependence, the mean shock radii of the simulations
\mesavlr{} and \mesavlroct{} reach $\sim$10\,km further out at
$\sim$100\,ms after bounce and then recede more slowly such that after
$\sim$250\,ms they are $ \sim$30\,km (or $\sim$ 30\%) further out
compared to \mesalr{} and \mesalroct{}. This larger shock radius
coincides with an increased PNS radius.  Since such a dramatic
difference in both the PNS radius and the shock radius is not seen in
our 1D models of this progenitor, we attribute the difference, as
least in part, to the presence of strong PNS convection which is
effectively supporting the PNS and allowing it to maintain a larger
radius and therefore a larger shock radius.  This effect, in relation
to its impact on the heavy-lepton neutrino luminosities, has been
noted in several recent multidimensional studies of CCSNe including
\cite{oconnor:2018, radice:2017}, but note that other multidimensional
simulations see a smaller impact \cite{buras:2006a}. We discuss this
further when we discuss the neutrino luminosities from our simulations
in \sref{sec:neutrinos}. Related to
this is the observation that in the one full 3D simulation with
velocity dependence we see no presence of the SASI.  In part, we
attribute this to the large shock radius discussed above.  A
consequence, as can be seen from \fref{fig:vdep}, is that after
200\,ms after bounce there is a larger ($\gtrsim $30\% more) turbulent
kinetic energy present in the gain region of \mesavlr{} when compared
to \mesalr{}. This can lead to a suppression of the growth of the
SASI \citep[cf.,][]{foglizzo:2007, fernandez:2014}.

\subsubsection{Impact of Dimensionality}

For completeness, we examine the dimensional dependence of this
progenitor model.  We perform a 1D, 2D (with and without progenitor
perturbations), and include our two 3D simulations with and without
progenitor perturbations in this comparison. Our 2D simulations do
successfully explode, but only after a post-bounce time of $\sim$1\,s
and $\sim$1.4\,s for \mesatwodpert{} and \mesatwod{}, respectively.
We note several interesting effects. First, it is clear that numerical
perturbations from the computational grid allow for the earlier growth
of lateral kinetic energy which acts, at least initially, to support a
larger shock radius. For the two 3D simulations, this growth starts at
$\sim$70\,ms after core bounce, the same time as the mean shock radius
deviates from the 1D result. In 2D, this is delayed until
$\sim$90\,ms.  The difference is due to the Cartesian grid,
particularly the prescribed mesh refinement boundaries, which trigger
convective growth. Differences in the block sizes, and the added
dimension, cause these numerical perturbations to trigger at different times.  Soon
after, as early as $\sim$140\,ms after bounce, the 2D lateral kinetic
energies, show a dramatic growth beyond the 3D equivalent.  In part,
this is expected because the nature of turbulence in 2D, which tends
to drive kinetic energy to large scales.  This alone does not explain
the increased values we see, however, when coupled in a CCSN
simulation, these large scale features also tend to drive shock
expansion, which has a non-linear, but positive effect for the
neutrino mechanism, i.e. increased neutrino heating and a greater
gain-region mass. 

As was the case in 3D, the 2D simulation with progenitor perturbations
are qualitatively and quantitively closer to explosion, especially
during the epoch when the perturbations from the silicon shell are
accreting through the shock. At this time, there is a larger mean
shock radius, larger neutrino heating, more lateral kinetic energy and
a higher $\tau_\mathrm{adv}/\tau_\mathrm{heat}$. As a cautionary note,
2D simulations of CCSNe can be very stochastic in nature and large
deviations in the evolution (even qualitative ones) can occur for
small deviations in the initial conditions.

\subsection{SASI and Angular Momentum}
\label{sec:SASI}

\begin{figure*}[th]
  \plotone{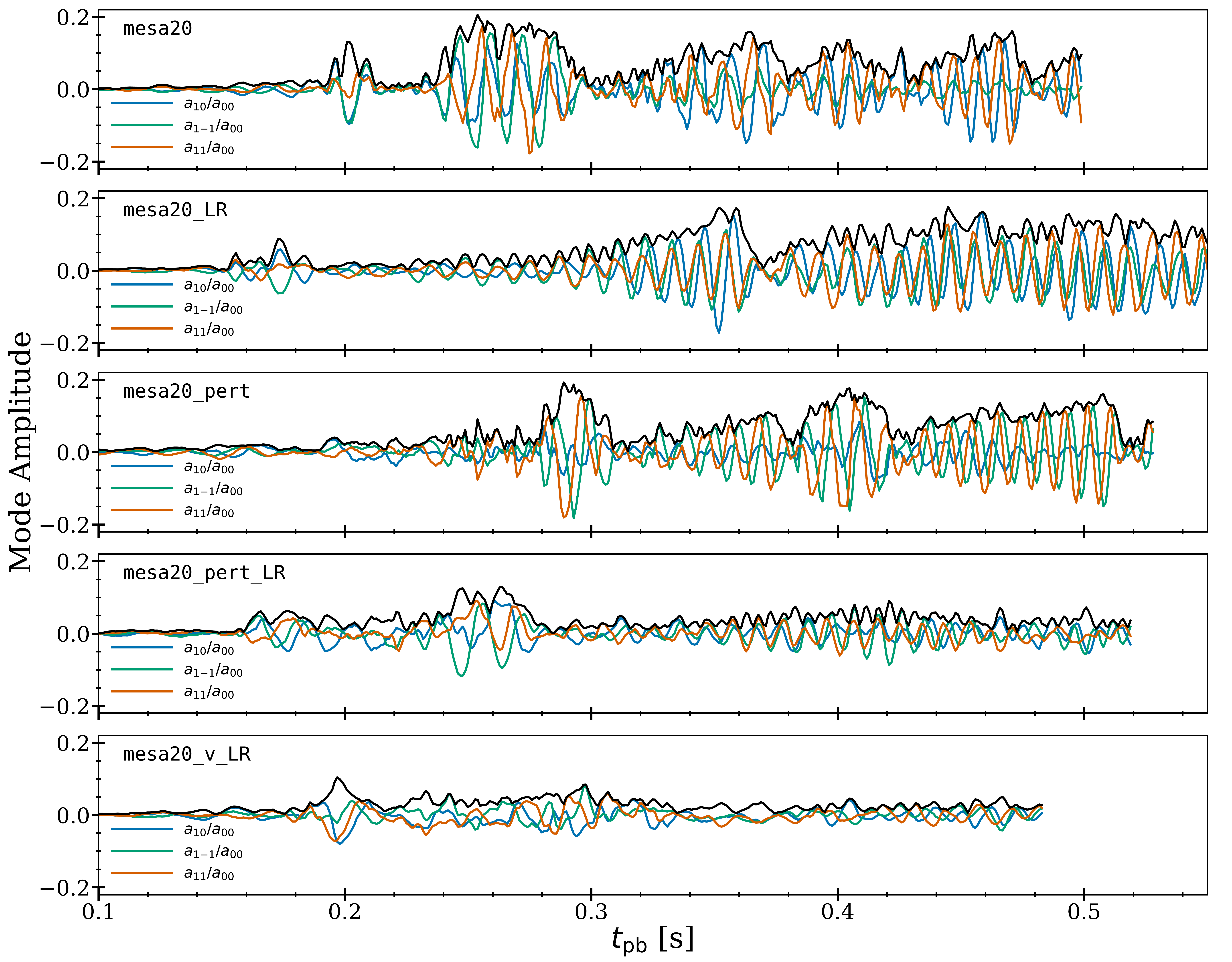}
  \caption{Spherical Harmonic decomposition of the shock in the
    \mesa{}, \mesalr{}, \mesapert{}, \mesalrpert{}, and \mesavlr{}
    simulations, from top the bottom, respectively. We show the
    individual $\ell=1$ components of the shock decomposition,
    relative to the $\ell=0$ (mean shock radius) value,
    $a_{10}/a_{00}$ (blue), $a_{1{-1}}/a_{00}$ (green),
    $a_{11}/a_{00}$ (orange), which correspond to
    $\langle y_\mathrm{sh} \rangle$, $\langle z_\mathrm{sh} \rangle$,
    and $\langle x_\mathrm{sh} \rangle$, respectively. Additionally,
    we show the square root of the total relative power in the
    $\ell=1$ mode compared to the $\ell=0$ mode as the solid black
    line.}\label{fig:mesa20_sasi}
\end{figure*}

We see characteristic signs of the standing accretion shock
instability (SASI) in most of our full 3D models. Using the spherical
harmonic decomposition analysis presented in \cite{couch:2014}, we
compute the spherical harmonic components of the shock
radius decomposition as a function of time for all of the full 3D simulations,
\mesa{}, \mesalr{}, \mesapert{}, \mesalrpert{}, and \mesavlr{}. Based
on the shock decomposition, the SASI
is clearly present in models \mesa{}, \mesalr{}, and \mesapert{}, to a
lesser extent in model \mesalrpert{}, and not discernable in model
\mesavlr{}. In \fref{fig:mesa20_sasi}, we show
the individual components of the $\ell=1$ spherical harmonic,
$a_{1\,-1}/a_{0\,0} = \langle z_\mathrm{sh} \rangle/\langle
r_\mathrm{sh}\rangle$,
$a_{1\,0}/a_{0\,0} = \langle y_\mathrm{sh} \rangle/\langle
r_\mathrm{sh}\rangle$, and
$a_{1\,1}/a_{0\,0} = \langle x_\mathrm{sh} \rangle/ \langle
r_\mathrm{sh}\rangle$, where $\langle x_{i,\mathrm{sh}} \rangle$
denotes the mean value of the Cartesian coordinate $x_i = \{x,y,z\}$
of the shock front and $\langle r_\mathrm{sh} \rangle$ is the mean
radius of the shock front. We also show the total relative power in the
$\ell=1$ mode, $|\bar{a}_{1}| \sim \sqrt{\sum_ma_{1\,m}^2}/a_{0\,0}$
as the solid black line. The peak amplitudes reach $\sim$0.15-0.2 for
the strongest three models, and $\sim$0.05 in \mesalrpert{}.  Most of
the SASI dominated phases are predominantly spiral modes where one
expects a constant total power rather than for a sloshing mode where
one would expect an oscillatory total power (at twice the frequency).
At times there are clear, but small, oscillatory features on top of
the total power (at twice the spiral mode frequency) that are due to a
relatively smaller sloshing mode component. For example, from
$\sim$250-325\,ms in model \mesalr{}, as discussed in detail below, as a
quintessential example.

\begin{figure}[tb]
  \plotone{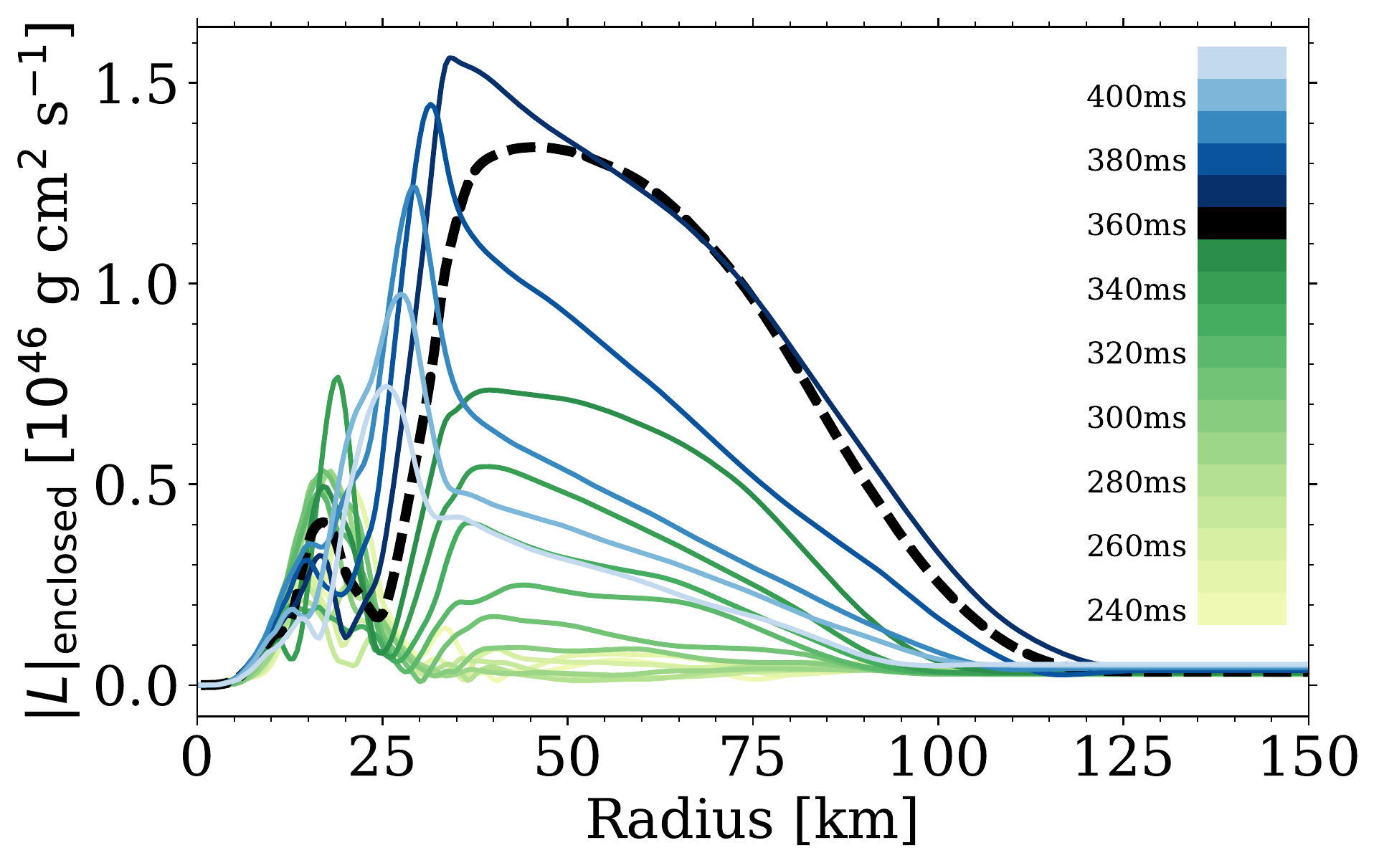}
  \caption{Radial profiles of the enclosed angular momentum, $|L|$,
    for 18 postbounce times between 240\,ms and 410\,ms in the
    \mesalr{} simulation. The dashed black lines denote the time when
    the SASI $\ell=1$ mode has the highest amplitude,
    $\sim$360\,ms. Green shades show the distribution as the SASI
    spiral mode is building up, while blue shades show distribution
    after the spiral SASI model is disrupted and the distributed
    angular momentum is isotropized near the PNS
    core.}\label{fig:mesa20_Lr_L}
\end{figure}

Spiral SASI modes, like the one seen from $\sim$250-360\,ms in model
\mesalr{}, redistribute angular momentum so that the PNS is
counter-rotating with the spiral mode, even in zero net-angular
momentum configurations \citep{Blondin:2007, Blondin:2007b,
  guilet:2014, Hanke:2013}. We observe this redistribution in our
models. In \fref{fig:mesa20_Lr_L}, we show the enclosed angular
momentum as a function radius for 18 times between 240\,ms and
410\,ms, in 5\,ms increments. At all times, there is net angular
momentum inside the PNS (from $\sim$10-25\,km) due to PNS
convection. However, for the most part, this cancels itself out near
the edge of the PNS convection zone. Outside of $\sim$25\,km, at early
times (yellows and light greens) there is little net enclosed angular
momentum. As the SASI spiral mode grows, starting around
$\sim$280\,ms, we begin to see the redistribution of this angular
momentum (dark greens towards black).  By $\sim$360\,ms (black dashed
line), where there is a peak of the $\ell=1$ power, $|\bar{a}_{1}|$,
the angular momentum redistribution is reaching a maximum.  Inside
$\sim 40$\,km, the PNS is counter-rotating against the spiral mode
with an enclosed angular momentum of
$\sim 1.5\times10^{45}$g\,cm$^2$\,s$^{-1}$. There is an equal, but
opposite amount trapped in the SASI spiral wave outside of
$\sim$40\,km, so that the net angular momentum of the system is
zero. After $\sim$360\,ms, we see the spiral SASI rapidly disrupts.
The angular momentum in the gain region accretes onto the PNS rather
than being sustained in the gain region.  This process takes place
over an accretion timescale $M_\mathrm{gain}/\dot{M} \sim 30$\,ms. The
amount of kinetic energy that is ultimately dissipated at the surface
of the PNS during this time is $\sim 10^{48}$\,ergs. Even if this was
to be completely transformed into heat over $\sim$30\,ms, the
additional heating would be much less than the steady-state neutrino
heating at this time.

\begin{figure}[tb]
  \plotone{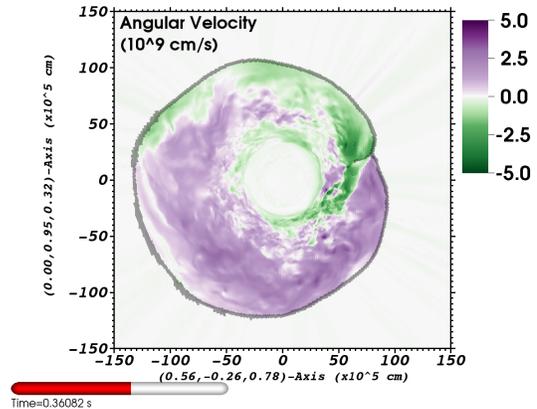}
  \caption{Slice through the plane aligned with the plane of the SASI
    spiral mode in the \mesalr{} simulation between the postbounce
    times of $\sim$240\,ms-$\sim$370\,ms. The colors denote the
    angular velocity ($v_\phi^\mathrm{SASI}$) of the material in the
    plane , purple shave are rotating counter-clockwise while green
    shades are rotating clockwise.  The direction of the spiral SASI
    wave is counter-clockwise. Zones tagged as shocks are shown as
    transparent gray.  (Animation available online)}\label{fig:spiralmodemotion}
\end{figure}

Complementary to \fref{fig:mesa20_Lr_L}, we show slices along the SASI
plane in \fref{fig:spiralmodemotion} (animation available online
showing the spiral mode build up and then the destruction;
otherwise we show the frame at $\sim$360\,ms) of the angular velocity
of the material in this plane.  Positive values (purples) are rotating
counter-clockwise while negative values (greens) are rotating
clockwise. The spiral SASI wave is rotating counter-clockwise, the
triple point is located near the right side of the figure. Zones that
are tagged as shocks are shown as transparent grey.  In the core we
see the build up of clockwise rotating material (green) on the surface of the
PNS, while the bulk of the gain region is rotating counter-clockwise (purple).

Just prior to $\sim$360\,ms when the spiral SASI mode peaks in
amplitude, at around $\sim$350\,ms, the shock starts a rapid expansion
(see \fref{fig:mesa20vspert}).  This is coincident with an increase in
the lateral kinetic energy and an increase in the neutrino heating,
likely as a result of the shock expansion.  After this time the spiral
SASI wave is disrupted.  These dynamics suggest that the SASI mode
becomes highly non-linear, disrupts, and subsequently leads to a
transient reenergization of the shock. This is quantitatively
demonstrated via the increase in the ratio
$\tau_\mathrm{adv}/\tau_\mathrm{heat}$. Clearly, this transient
reenergization is not enough to launch as explosion, but it shows a
potential way for the SASI to lead to an explosion.   This phenomena is seen in
other simulations present in this work (c.f.  the peaks of the SASI
mode in \mesa{} at $\sim$400\,ms and $\sim$470\,ms, and \mesapert{} at
$\sim$370\,ms and $\sim$510\,ms). It is important to note that not all
transient shock expansions seen in \fref{fig:mesa20vspert} are
followed by a collapsed SASI wave, nor is it the case that all SASI
wave collapses follow a rapid shock expansion, yet there does seem to
be a high correlation between these phenomena.

Provided an explosion was launched during such an event, it may well
be the case that the PNS retains the distributed angular momentum
after the explosion as suggested in \citep{Blondin:2007,
  Blondin:2007b, guilet:2014}. Assuming the neutron star ends up with
a total angular momentum of
$1.5\times 10^{46}$\,g\,cm$^{2}$\,s$^{-1}$, and a gravitational mass
of $\sim$1.6$M_\odot$ (based on the baryonic mass of the PNS at
360\,ms of $\sim$1.8$M_\odot$ and the SFHo EOS) and therefore a moment
of inertia of $\sim 1.7\times10^{45}$\,g\,cm$^2$, the final period
would be $\omega = 8.8$\,rad\,s$^{-1}$, or $\sim$700\,ms. This is
roughly a factor of 5 faster than the prediction of \cite{guilet:2014},
\begin{eqnarray}
\nonumber
P &\sim &290I_\mathrm{45}\left(\frac{10}{\kappa}\right)
  \left(\frac{P_\mathrm{SASI}}{50\,\mathrm{ms}}\right)
  \left(\frac{120\,\mathrm{km}}{r_\mathrm{sh} - r_*}\right)
  \left(\frac{v_\mathrm{sh}}{3000\,\mathrm{km}\,\mathrm{s}^{-1}}\right)\\
  &&\left(\frac{0.3M_\odot
    \mathrm{s}^{-1}}{\dot{M}}\right)\left(\frac{150\,\mathrm{km}}{r_\mathrm{sh}}\right)^2\left(\frac{r_\mathrm{sh}}{3\Delta
  r}\right)^2\,\mathrm{ms}
\end{eqnarray}
where for our case, $\kappa\sim 9$, $P_\mathrm{SASI} \sim 13$\,ms,
$r_\mathrm{sh} \sim 110$\,km, $r_* \sim 40$\,km,
$v_\mathrm{sh} \sim 8000$\,km\,s$^{-1}$, $M_\odot \sim 0.3$, and
$\Delta r / r_\mathrm{sh} \sim \sqrt{2}|\bar{a}_{1}| \sim 0.2$ leading
to a prediction of $P \sim 3500$\,ms. This difference could be due to
any number of differences between the idealized models of
\cite{guilet:2014} and the work here. In particular, our SASI spiral
wave is confined to a much smaller radius as the mean shock radius is
only 110\,km, compared to the typical value of 150\,km in
\cite{guilet:2014}.  Also, the more massive PNS (1.8\,$M_\odot$), and
the smaller mean shock radius, gives a larger typical radial velocity
behind the shock (8000\,km\,s$^{-1}$ compared to
3000\,km\,s$^{-1}$. There could be non-linear effects associated with
the large deviations from the assumed values and scalings in
\cite{guilet:2014}.

\subsection{Neutrinos}
\label{sec:neutrinos}

\begin{figure*}[tb]
  \plotone{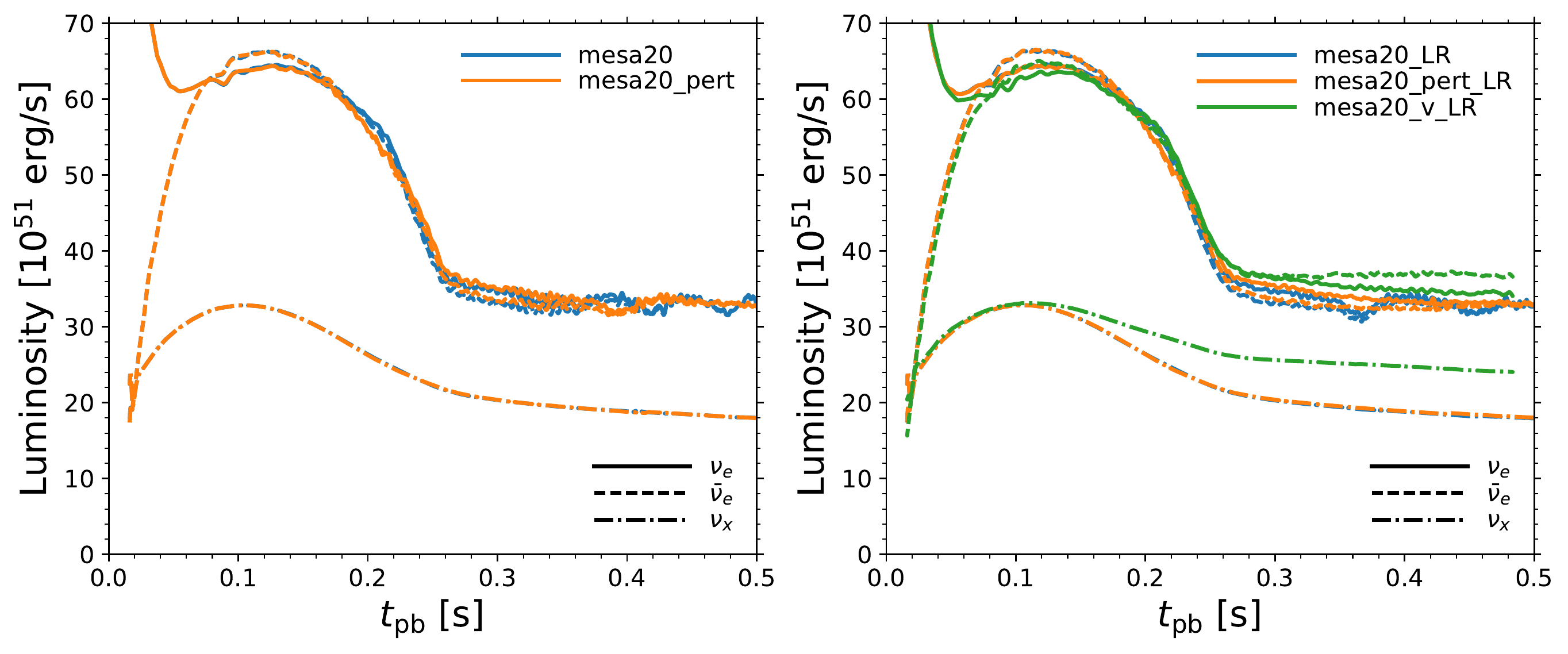}
  \caption{Sky-average neutrino luminosities for all three species ($\nu_e$,
    $\bar{\nu}_e$, and a single $\nu_x$) as a function of postbounce time for
    the five, full-3D models explored in this article. In the left
    panel we show the luminosities from \mesa{} and \mesapert{}, while
    in the right panel we show the luminosities from \mesalr{},
    \mesalrpert{}, and \mesavlr{}.}\label{fig:lums}
\end{figure*}

We show the neutrino luminosities produced by each one of our five
full-3D simulations in \fref{fig:lums}.  For clarity, we divide them
into our standard resolution (left) and the low resolution
(right). The solid lines show the electron neutrino luminosity, the
dashed lines denote electron antineutrino luminosity, and the
dashed-dotted lines show the luminosity of a single heavy-lepton
neutrino species. The simulations being at 15\,ms after bounce.  In
the beginning of the non-velocity dependent simulations there is a
sharp spike in the luminosities due to the transition from the
velocity-dependent transport in GR1D and the transport in FLASH.  This
spike is much smaller in the velocity-dependent FLASH simulations.
Since the underlying progenitor is the same for each simulation we do
not expect large variations in the predict neutrino signals between
simulations.  That being said, we do see several noteworthy
differences that we will explicitly mention. As the system evolves we
see the rise of both the $\bar{\nu}_e$ and $\nu_x$ luminosities and
the fall of the $\nu_e$ luminosity as the remaining iron core accretes
on to the PNS.  The electron type luminosities plateau at
$\sim$65\,$\times 10^{51}$\,erg\,s$^{-1}$ while the $\nu_x$ luminosity
plateaus at $\sim$33\,$\times 10^{51}$\,erg\,s$^{-1}$.  At a
postbounce time of $\sim$220\,ms we begin the dramatic drop in
luminosity that is a result of the mass accretion rate drop associated
with the accretion of the silicon/oxygen interface through the shock.
We do notice a difference between the simulations with and without the
imposed progenitor perturbations.  The \mesapert{} and \mesalrpert{}
simulations (shown in orange) have a shallower drop in luminosity.  It
starts earlier and ends later.  This is not due to differences in the
mass accretion rate onto the shock, but rather due to a longer
advection time through the gain region and therefore a slower (for a
short time) mass accretion rate into the cooling region.  The longer
advection time is due to the additional support provided to the shock
by the turbulence motions seeded by the progenitor perturbations.  We
see a similar phenomena at later times, when collapsing spiral SASI
waves trigger increased turbulence activity (see \sref{sec:SASI}),
increased advection time, and small drops in the
neutrino luminosity.  For example, at $\sim$370\,ms in \mesalr{},
$\sim$400\,ms in \mesapert{}, and $\sim$410\,ms and $\sim$480\,ms in
\mesa{}.

The luminosities presented in \fref{fig:lums} are spherically
averaged.  However, since our transport code is multidimensional, we
can easily examine the variation of the neutrino signal over the
emission direction.  We perform the same spherical harmonic
decomposition on the individual neutrino signals (extracted at
500\,km) as we applied on the shock surface. However, for neutrinos it
is convient to consider the decomposition in the form of
\citep{Tamborra:2014, melson:2016}
\begin{equation} 
L_\nu \sim L_\nu^\mathrm{monopole} + L_\nu^\mathrm{dipole}\cos{\theta}\,,
\end{equation}
where the monopole component is related to the $\ell=0$ decomposition,
$L_\nu^\mathrm{monopole} = (a_{00}^2)^{1/2}$, the dipole component is
related to the $\ell=1$ decomposition,
$L_\nu^\mathrm{dipole} = 3\times (\sum_{i=-1}^1 a_{1i}^2)^{1/2}$, and
$\cos(\theta)$ is the cosine of the angle between the observer and the
direction of the dipole at any given time.  We do find a non-spherical
(i.e. dipole) component that is highly correlated with the SASI
motions. For $\nu_e$ and $\bar{\nu}_e$ we find
$L_{\nu_e/\bar{\nu}_e}^\mathrm{dipole} =
3|\bar{a}_1|_{\nu_e/\bar{\nu}_e} \sim 0.03 L_{\nu_e/\bar{\nu}_e}^\mathrm{monopole}$.  The $\nu_x$ signal also
has a directional variation.  However, it is roughly 4 times smaller,
i.e. $L_{\nu_x}^\mathrm{dipole} = 3|\bar{a}_1|_{\nu_x} \sim 0.008 L_{\nu_x}^\mathrm{monopole}$. As
the spiral SASI wave rotates around, so to does the dipole component
of the neutrino luminosities.  Over the course of one SASI period, the
neutrino luminosities as seen by an observer viewing the spiral SASI
wave edge on would have a trough-peak variation of $\sim$6\%
($\sim$1.5\%) for $\nu_e$ and $\bar{\nu}_e$ ($\nu_x$).  This is
smaller than seen in \cite{tamborra:2013}, where peak-trough
variations of order $\sim$25\% were seen for some spiral SASI waves.
The phase of the neutrino modulation is interesting.  After taking
into account the light travel time from the PNS where the neutrinos
are emitted and at 500\,km where they are measured, we find that the
$\nu_e$ and $\bar{\nu}_e$ luminosities peak in the direction opposite
the average shock position, i.e. they lag the direction determined by
the $\langle x_{i,\mathrm{sh}} \rangle$ by 180$^\circ$. The $\nu_x$
luminosities generally peak at a different time then the $\nu_e$ and
$\bar{\nu}_e$ neutrinos. We attribute this to a variation in the
location of the peak values of the density and temperature within the
spiral SASI plane as a function of radius. The electron type neutrinos
and antineutrinos are emitted at a much larger radius than the
heavy-lepton neutrinos.  The accretion waves take longer to reach the
deeper radii where the heavy-lepton neutrinos are emitted. The exact
phase difference between the neutrino types smoothly varies with time.

Finally, we remark on the largest difference seen in \fref{fig:lums}.
As discussed in \sref{sec:vdep}, when we include velocity dependence
into our simulations, i.e. in the \mesavlr{} and \mesavlroct{}
simulations, we see an increased level of PNS convection. This
dramatically increases (by upwards of $\sim$30\% compared to the
non-velocity dependent simulations) the heavy-lepton neutrino
luminosity starting from $\sim$100\,ms after bounce.  The convection
brings hotter material from deeper down to larger radii where emitted
neutrinos can more easily escape.  This also increases the radii of
the heavy-lepton emission region, giving a larger effective area and
therefore a larger luminosity. As mentioned in \sref{sec:vdep}, this
effect has been noted in several recent multidimensional studies of
CCSNe including \cite{oconnor:2018, radice:2017}.  There is tension
regarding the total expected enhancement due to multidimensional
effects as we note that other multidimensional simulations see a
smaller impact \citep{buras:2006a}.

\subsection{Lepton-number Emission Self-sustained Asymmetry: LESA}
\label{sec:lesa}
The Lepton-number Emission Self-sustained Asymmetry, or LESA,
phenomena was first reported in \cite{Tamborra:2014}.  The LESA is
characterized by a spatial asymmetry in the total electron number
emission via a strong dipole component. The LESA can be described with
a monopole and dipole component,
\begin{equation}
N_{\nu_e} - N_{\bar{\nu}_e}   = A_\mathrm{monopole} + 
A_\mathrm{dipole}\cos{\theta}
\end{equation}
where $\cos(\theta) = \hat{r} \cdot \hat{r}^\nu_\mathrm{dipole}$.  In
some cases reported in \cite{Tamborra:2014}, the dipole component was
comparable to or larger than the monopole component.  The consequence
is that in the direction of $-\hat{r}^\nu_\mathrm{dipole}$, the net
lepton number emission was negative, i.e. more electron
antineutrinos were emitted then electron neutrinos.  Such a situation
many have dramatic consequences for neutrino oscillations, neutrino
detection (parameter inferences), and nucleosynthesis, to name a few.

Since \cite{Tamborra:2014}, the LESA has been reported in all 3D
models from the Garching collaboration using the PROMETHEUS-VERTEX
code \citep{janka:2016} but has never been conclusively demonstrated
by an independent source.  In addition to being only observed by one
simulation code, another large criticism of the LESA instability is
that the neutrino transport scheme employed made use of the ray-by-ray
approximation. Our neutrino transport scheme is completely independent
of PROMETHEUS-VERTEX. Importantly, it does not make the ray-by-ray
approximation but rather solves the multidimensional transport of the
neutrinos directly.  For these reasons, a search for LESA in our
simulations is warranted.

\begin{figure*}[tb]
  \plottwo{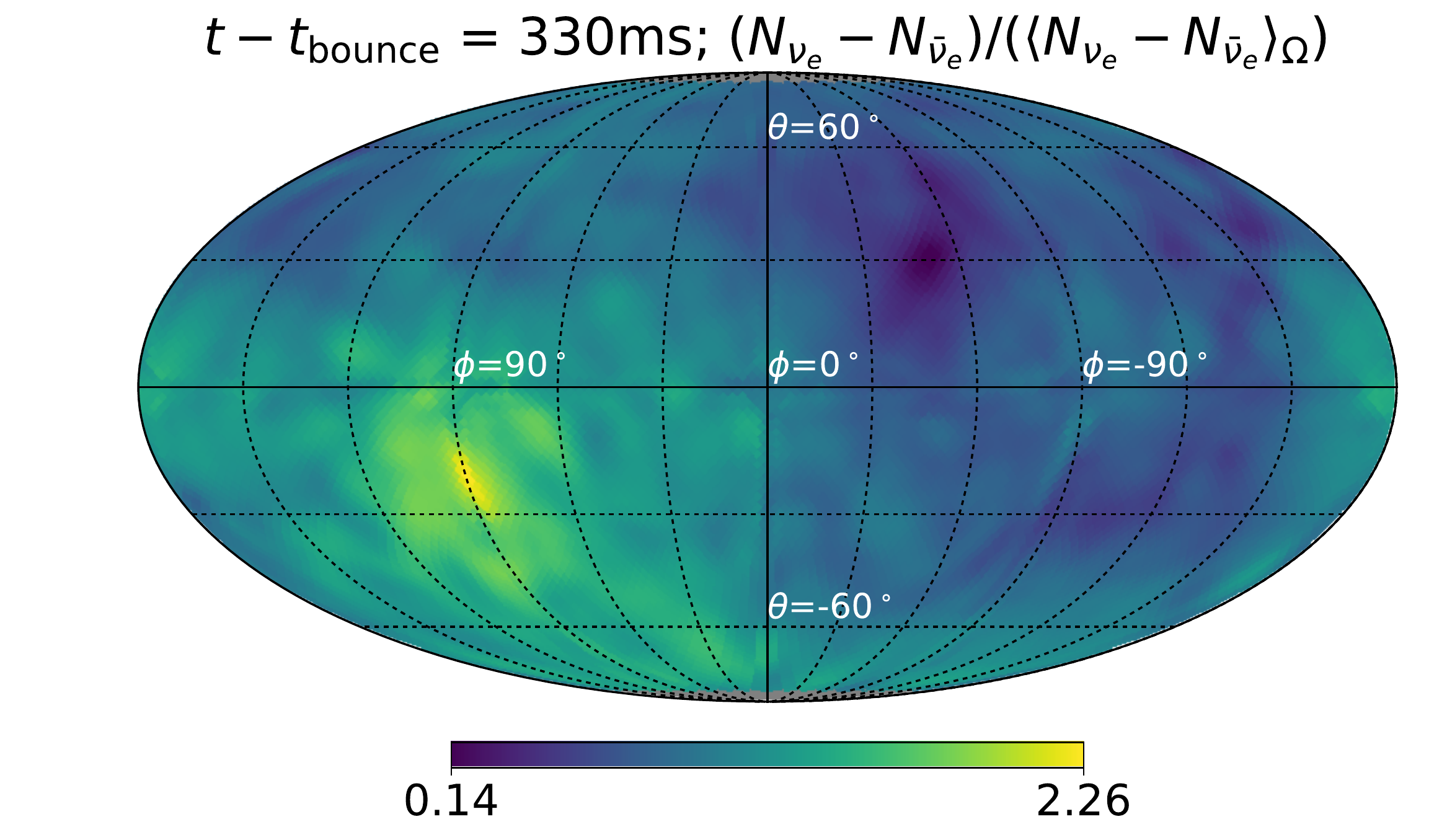}{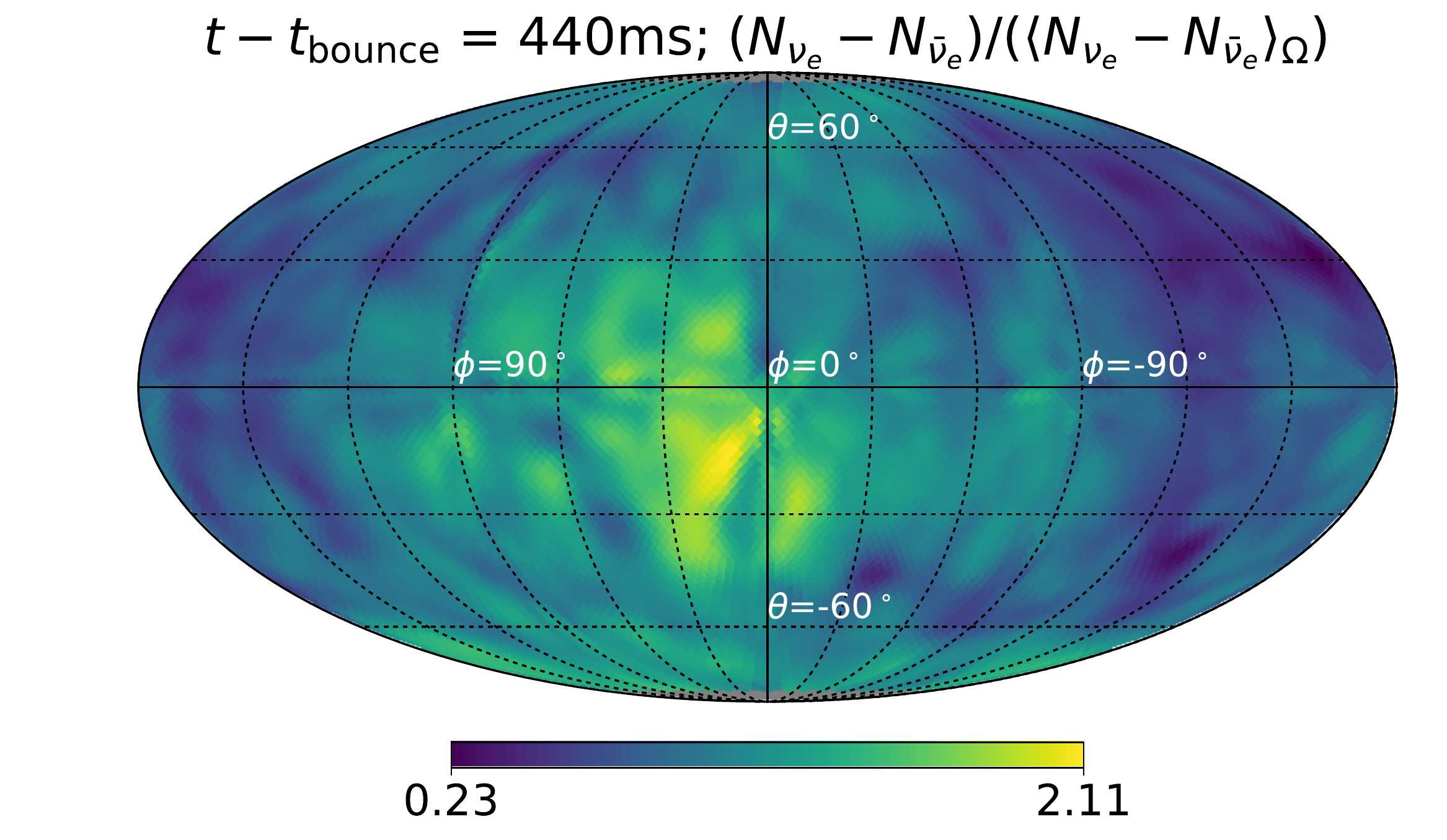}
  \caption{Relative net lepton number flux passing through 500\,km for
    two snapshots of the \mesavlr{} simulations.  A value of 1 in
    these figures represents the spherical average of net lepton flux,
    positive values denote an excess of positive lepton number
    (i.e. more electron neutrinos) and negative numbers denote a
    an excess of negative lepton number (i.e. more electron
    antineutrinos). Since the protoneutron star is deleptonizing, we
    expect a significant excess of lepton number.}\label{fig:LESAskymaps}
\end{figure*}

\begin{figure}[tb]
  \plotone{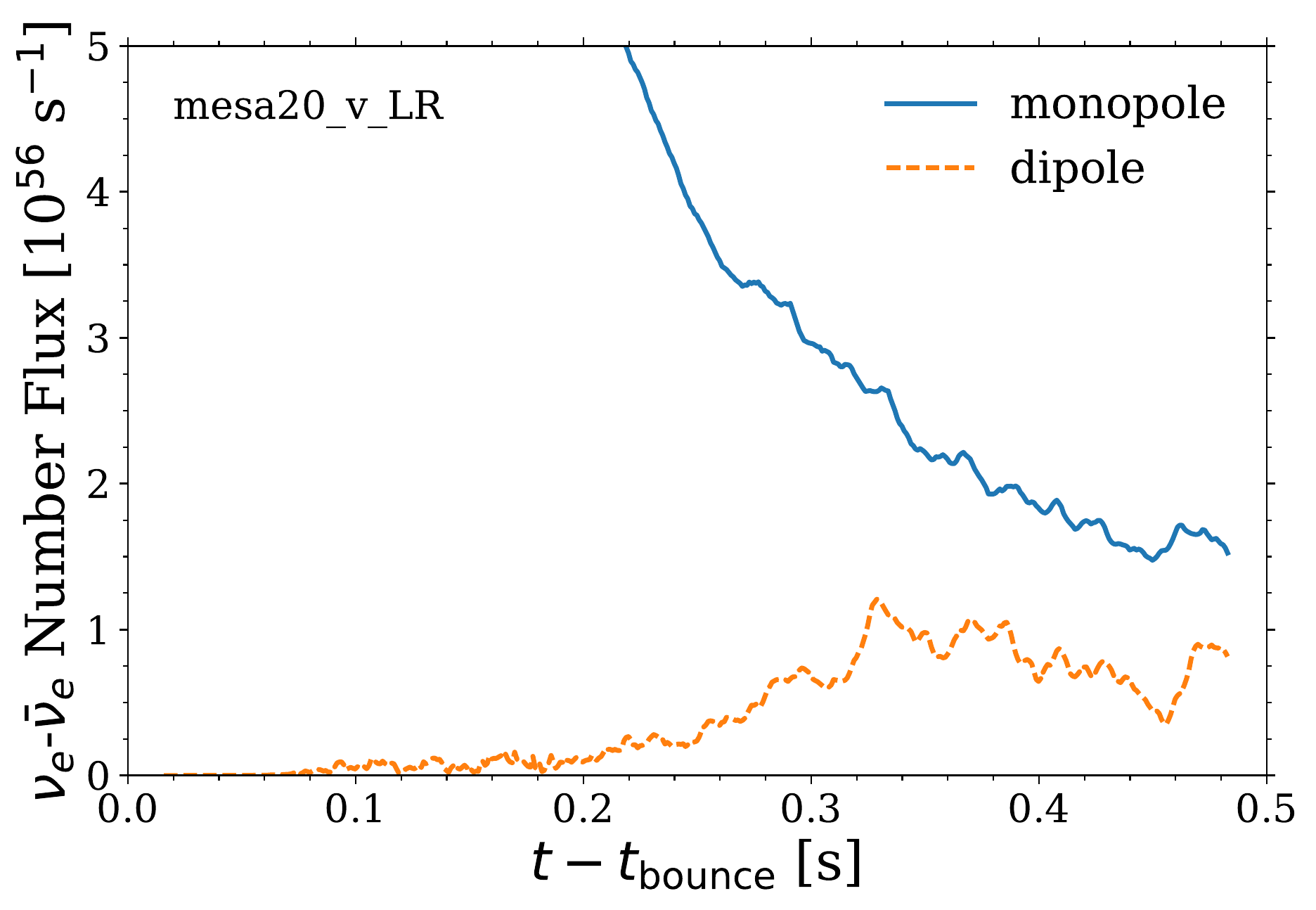}
  \plotone{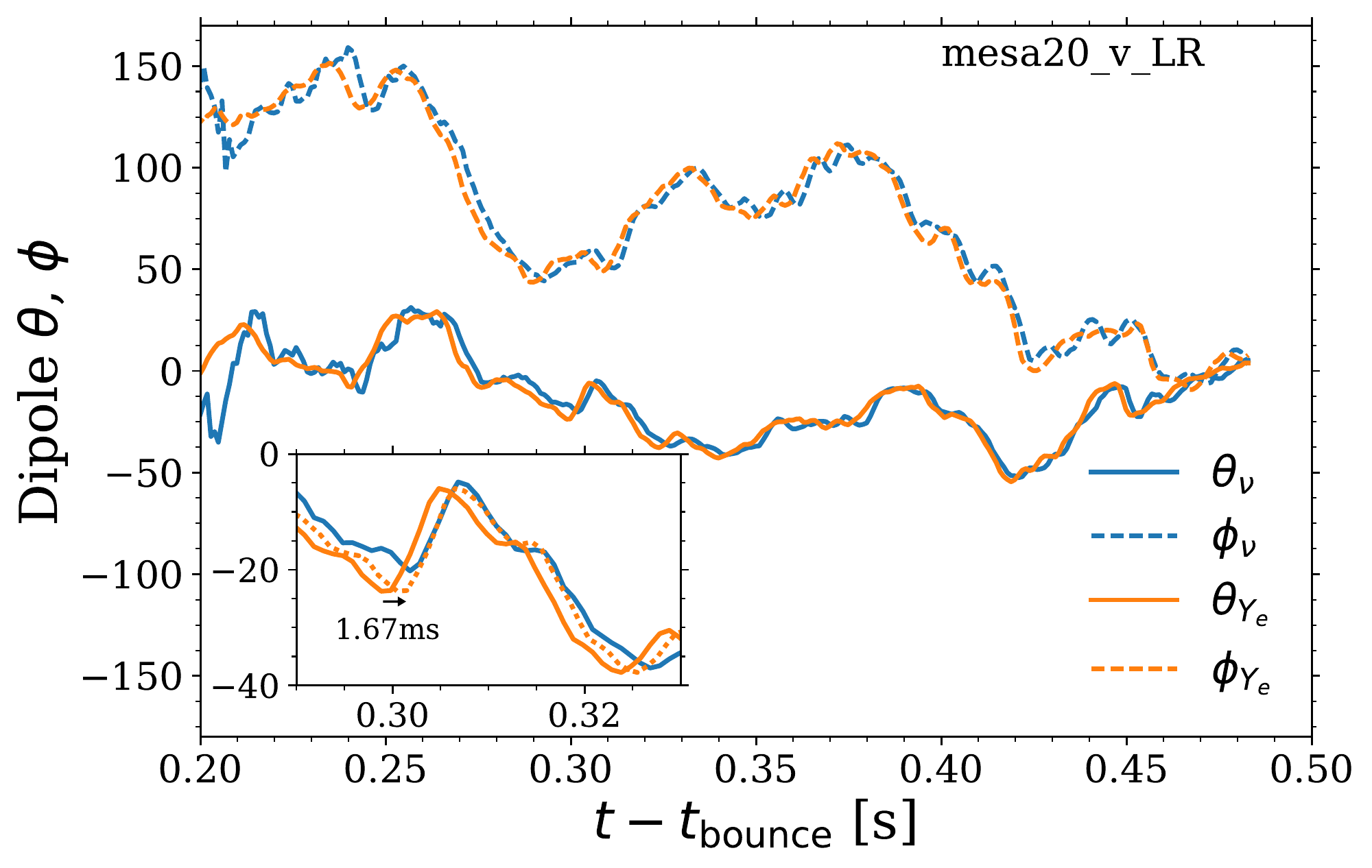}
  \caption{LESA properties for model \mesavlr{}. We show the first two
    components of the lepton number emission distribution as a
    function of postbounce time (top panel), and the evolution of the
    dipole component directions (bottom panel). In the bottom panel we
    also show the direction of a $Y_e$ dipole,
    $\hat{r}^\mathrm{Y_e}_\mathrm{dipole}$, as defined in
    \eref{eq:yedipole}, and an inset showing more clearly the slight
    time lag, roughly equivalent to the light travel time of the
    neutrinos to the 500\,km sphere where they are
    measured.}\label{fig:LESAovertime}
\end{figure}

We search our simulations for evidence of LESA.  We extract from our
simulations the net lepton number flux through a sphere with radius
500\,km.  In \fref{fig:LESAskymaps}, we show the relative net lepton
flux
$(N_{\nu_e} - N_{\bar{\nu}_e}) / \langle N_{\nu_e} - N_{\bar{\nu}_e}
\rangle_\Omega $ from the \mesavlr{} simulation for two times (330\,ms
and 440\,ms after bounce). The presence of a dipole is clear, located
at $\phi \sim 90^\circ (\sim 15^\circ)$ and
$\theta \sim-20^\circ ( \sim -15^\circ)$ for $t_\mathrm{pb} = 330$\,ms
(440\,ms). At 1\,ms time intervals, we decompose the net lepton number
flux ($N_{\nu_e} - N_{\bar{\nu}_e}$) sky map into spherical harmonics.
Recall from \sref{sec:neutrinos}, this decomposition gives both
$A_\mathrm{monopole} = (a_{00}^2)^{1/2}$ and
$A_\mathrm{dipole} = 3\times (\sum_{i=-1}^1 a_{1i}^2)^{1/2}$, where
$a_{lm}$ are the spherical harmonic decomposition components
\citep{couch:2014}. This notation is such that if the monopole and
dipole term are equal in amplitude (and the dipole term is positive),
then an observer located along the dipole direction sees twice the
average net lepton number flux and an observer along the anti-dipole
direction sees zero net lepton number flux.  In model \mesavlr{} we
see a strong presence of LESA, as expected based on the sky maps seen
in \fref{fig:LESAskymaps}. In the top panel of
\fref{fig:LESAovertime}, we show $A_\mathrm{monopole}$ and
$A_\mathrm{dipole}$ for model \mesavlr{}. The dipole component begins
to grow at $\sim$200\,ms after bounce and reaches an appreciable
fraction of the monopole component (at times $\sim$50\%) by
$\sim$300-400\,ms after bounce.  In addition to having a large ratio
of the dipole to monopole component, we also see that the LESA is
stable (but migratory) in its position.  In the bottom panel of
\fref{fig:LESAovertime}, we show, in blue, the $\theta$ (solid) and
$\phi$ (dashed) values of the dipole direction,
$\hat{r}^\nu_\mathrm{dipole}$, determined from the individual
spherical harmonic components $a_{1m}$ of the net lepton number
flux. Prior to 200\,ms the dipole strength is very low and therefore
the dipole direction is not well constrained. We exclude this time
from the figure for clarity. We also show the simultaneous position of
an asymmetry in the electron fraction near the PNS convection zone,
$\hat{r}^\mathrm{Y_e}_\mathrm{dipole}$. We evaluate the direction of
this dipole between the radii of 25 and 30\,km (where the protoneutron
star convection is strongest) as
\begin{equation}
\hat{r}_\mathrm{dipole}^\mathrm{Y_e} = \langle x_i Y_e \rangle /
(\sum_{j=1,2,3} \langle x_j Y_e \rangle^2)^{1/2}\,.\label{eq:yedipole}
\end{equation}
The direction of this dipole is remarkably similar to
$\hat{r}^\nu_\mathrm{dipole}$.  In fact, as shown in the inset of the
bottom panel in \fref{fig:LESAovertime}, the dipole direction in $Y_e$
slightly precedes the dipole direction in the net lepton number flux.
This lag time is roughly $\sim$2\,ms, and is due to the light travel
time between the location of the $Y_e$ dipole and the sphere at which
the neutrino number fluxes are measured.  To make this more clear, in
the inset we show both the true (solid orange line) and a shifted
version (dotted orange line; shifted by 500\,km/$c$ = 1.67\,ms) of the
$\hat{r}_\mathrm{dipole}^\mathrm{Y_e}$. The alignment of these dipoles
is not unexpected. The asymmetry in $Y_e$ is directly responsible for
the asymmetry in the net lepton number flux. The matter in the
direction of $\hat{r}_\mathrm{dipole}^\mathrm{Y_e}$ is on average more
electron rich and therefore emits relatively more electron neutrinos
compared to electron antineutrinos. Unlike the asymmetry in the net
lepton number, the asymmetry in the electron fraction can be seen
earlier than 200\,ms. However, its position is less stable at this
time. Although not shown, we also see a small dipole component to the
total $\nu_x$ flux that is anti-aligned with the dipole direction
(i.e. anti-aligned with the excess flux of the $\nu_e$s and aligned
with the excess flux of the $\bar{\nu}_e$s). This amplitude of this
$\nu_x$ dipole component is $\sim$25\% of the amplitude of the $\nu_e$
and $\bar{\nu}_e$ dipole components, which is in agreement with
\cite{Tamborra:2014}.

\begin{figure}[tb]
  \plotone{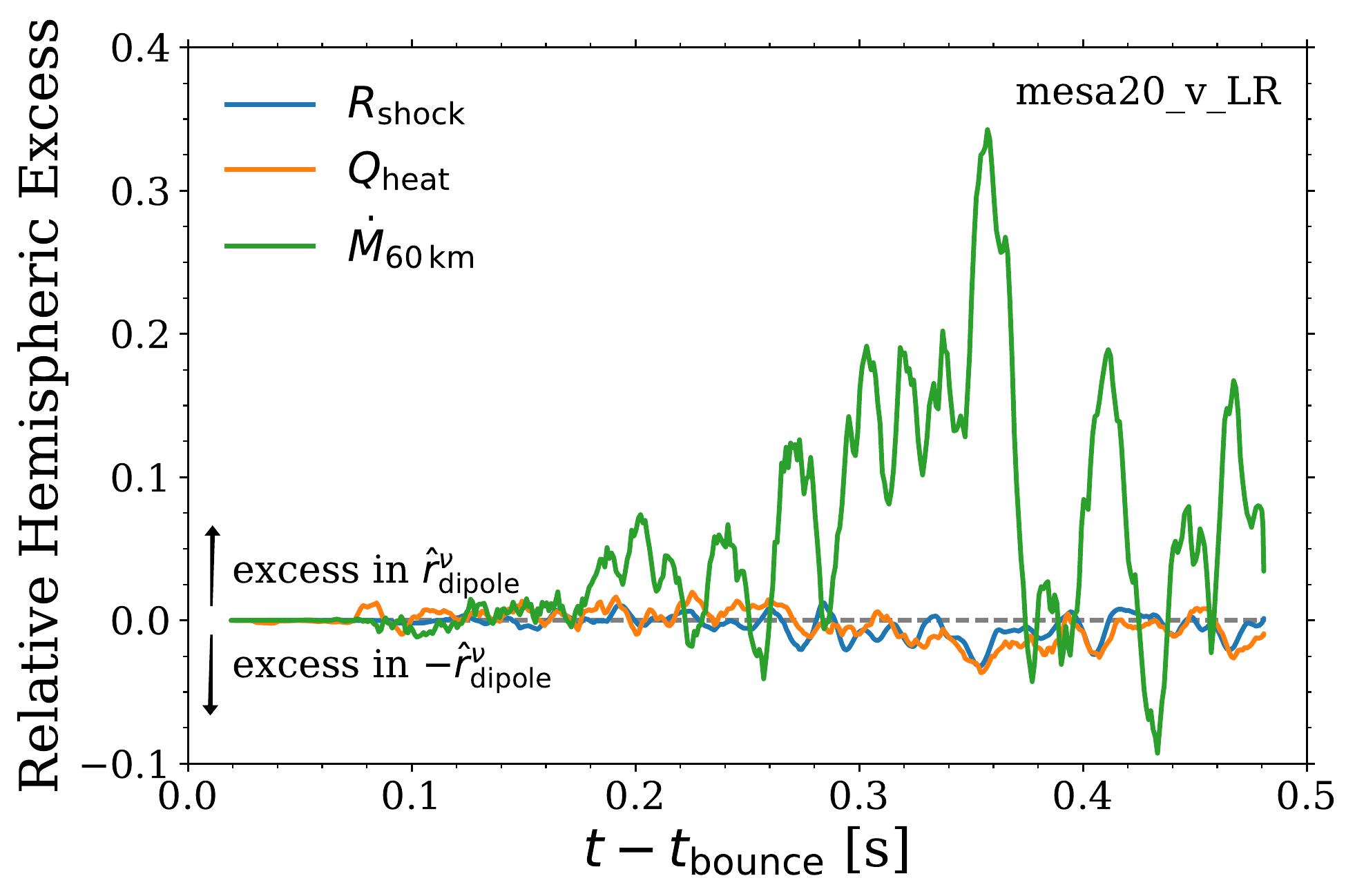}
  \caption{Relative difference between the average shock radius,
    neutrino heating, and the mass accretion rate on the protoneutron
    star as measured in the hemisphere aligned with the LESA dipole
    and as measured in the hemisphere anti-aligned with the LESA
    dipole.  According to the proposed mechanism from
    \cite{Tamborra:2014}, in the direction opposite the LESA dipole
    there is more neutrino heating and therefore a larger shock radius
  which funnel accretion into the hemisphere aligned with the LESA
  dipole.}\label{fig:LESAhemi}
\end{figure}

The cause of the asymmetry in $Y_e$ is not yet clear.
\cite{Tamborra:2014} propose a self-sustaining mechanism where the
increase of electron antineutrinos in the hemisphere of
$-\hat{r}^\nu_\mathrm{dipole}$ (and the decrease in the hemisphere of
$\hat{r}^\nu_\mathrm{dipole}$) leads to relatively stronger neutrino
heating, a larger shock radius, and consequently funneled accretion
onto the opposite hemisphere (i.e. near the direction of
$\hat{r}^\nu_\mathrm{dipole}$). This accretion of predominately
electron rich material sources the imbalance of the lepton emission
and, according to \cite{Tamborra:2014} excites the enhanced
protoneutron star convection.  Our simulations lend support to this
idea.  We show in \fref{fig:LESAhemi} the relative difference in the
shock radius, neutrino heating, and mass accretion rate (measured
between 55\,km and 65\,km) as measured in the two hemispheres defined
by the direction determined via the $Y_e$ asymmetry.  Specifically, we
show,
\begin{equation}
\Delta = \frac{X^+ - X^-}{\frac{1}{2} (X^+ + X^-)}
\end{equation}
where the $+$ denotes the hemisphere defined with
$\vec{r} \cdot \hat{r}_\mathrm{dipole}^\mathrm{Y_e} > 0$ and $-$
denotes the hemisphere where
$\vec{r} \cdot \hat{r}_\mathrm{dipole}^\mathrm{Y_e} < 0$. We see a
significant (upwards of $\sim$30\% at times) higher mass accretion
rate in the hemisphere aligned with
$\hat{r}_\mathrm{dipole}^\mathrm{Y_e}$, and therefore with
$\hat{r}_\mathrm{dipole}^\mathrm{\nu}$. Near the time when the LESA is
strongest (between $\sim$300\,ms and $\sim$450\,ms), we also see lower
average shock radii and neutrino heating rates (upwards of
$\sim$2-3\%) in the hemisphere aligned with
$\hat{r}_\mathrm{dipole}^\mathrm{Y_e}$.  The sign of each of these
relative ratios, agree with the observations of
\cite{Tamborra:2014}. Furthermore, for the most similar progenitor
used in \cite{Tamborra:2014}, s20, the magnitudes of these differences
roughly correspond to each other. Interestingly, the mass accretion
rate asymmetry begins to form early (before 200\,ms) before
significant asymmetry forms in the shock radius, neutrino heating, or
even net lepton number emission.  This suggest that although
asymmetric heating via a LESA dipole may funnel mass accretion into
the opposite hemisphere and thereby closing the self-sustaining cycle,
here may be other mechanisms that cause the mass accretion asymmetry
in the first place in order to initiate the growth of the LESA dipole.  

In the full 3D models without velocity dependence, (\mesa{},
\mesapert{}, \mesalr{}, and \mesalrpert{}) we do not see any
conclusive evidence for LESA. As discussed in \cite{Tamborra:2014} and
explored above, protoneutron star convection is clearly critical for
the LESA to develop as the neutrino lepton number asymmetry stems from
the asymmetry in the electron fraction distribution near and below the
neutrinospheres as shown above. We do not attribute the presence of
LESA in the \mesavlr{} simulations to the explicit inclusion of the
velocity dependence in the transport equations themselves, rather, we
suggest that it is due to the effect of the improved neutrino
transport on the protoneutron star convection.  With the included
velocity dependence, neutrino are able to advect with the flow.  We
see an earlier onset of protoneutron star convection in the \mesavlr{}
simulation (at $\sim$100\,ms after bounce) compared to, for example,
the \mesalr{} simulation (which does not have equivalent protoneutron
star convection until $\sim$200\,ms after bounce).  Furthermore, at a
given time, the protoneutron star convection is stronger (and the
region is wider) in the velocity dependent simulation compared to the
simulations without velocity dependence. Evidence for this conclusion
is also provided by an examination of the 3D models from
\cite{couch:2014}. In these simulations we see the presence of a
strong asymmetry in the electron fraction distribution in the
protoneutron star core and the expected effect on the neutrino
emission. Due to the low fidelity of the neutrino transport
(i.e. neutrino leakage), we find this inconclusive evidence for the
LESA. However, and important for this discussion, these models did
show signs of strong protoneutron star convection (even stronger than
seen in \mesavlr{}).

\subsection{Gravitational Waves}

\begin{figure*}[tb]
  \plotone{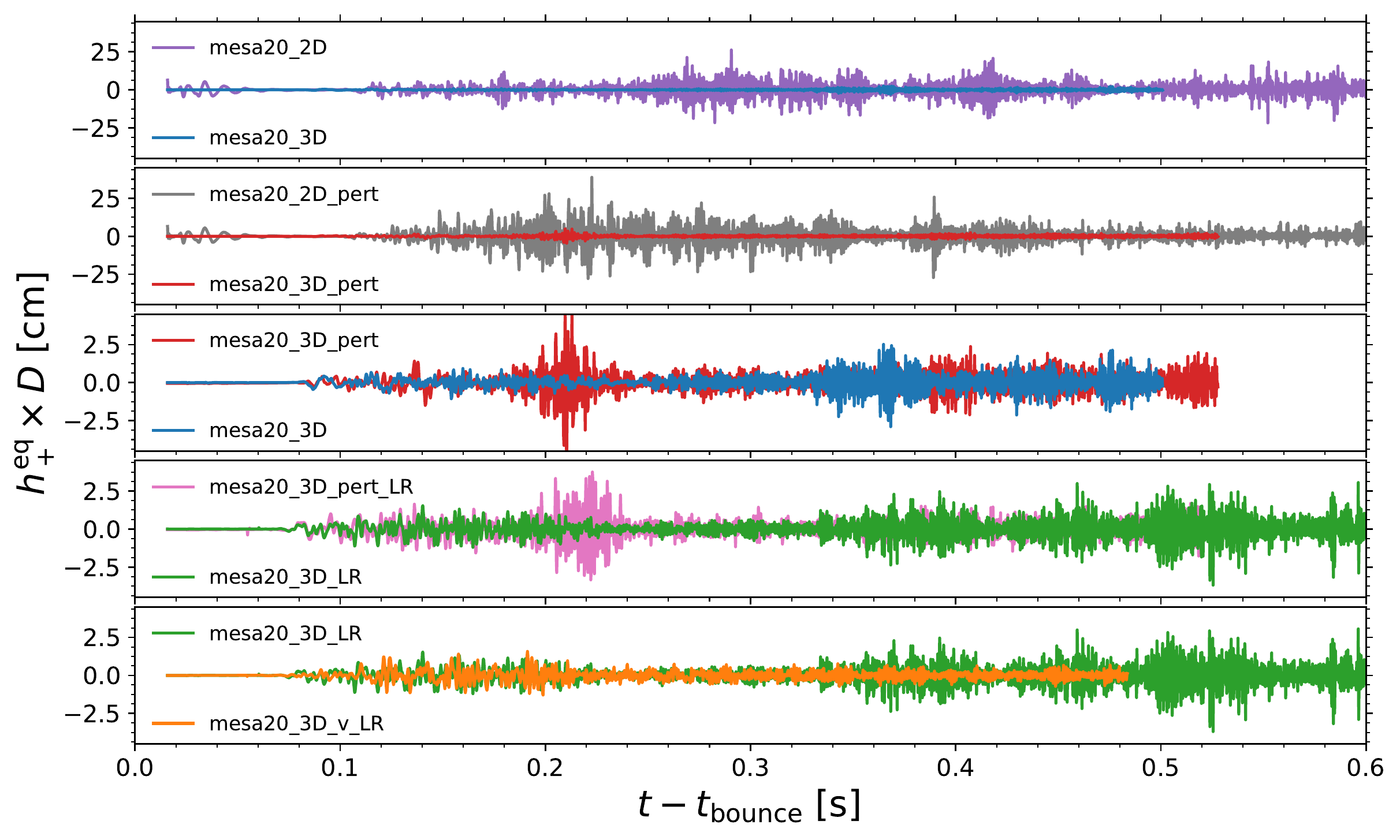}
  \caption{Gravitational wave signals ($h_+ \times D$) as measured by
    an observer located on the equator for various 2D and 3D
    simulatons.  This is the only non-zero signal in
    2D and is representative of the various possible 3D signals.  In
    the top two panels we compare 2D vs. 3D GWs, with and without
    perturbations. In the bottom three panels we compare 3D
    simulations: perturbations vs. no perturbations, standard
    resolution vs. low resolution, and velocity dependance vs. no
    velocity dependance. The small glitches in the 3D data visible
    at $\sim$50\,ms are due to the shock crossing the mesh refinement
     boundaries. The 2D GWs have a significantly higher amplitude than
    the 3D GWs.}\label{fig:gwtrains}
\end{figure*}

Using the quadrupole formula \citep{reisswig:2011a}, we estimate
the gravitational wave signal from our 2D and 3D simulations by
computing the first time derivative of the reduced mass-quadrupole
tensor via,
\begin{equation}
\frac{d I_{jk}}{dt} = \int  (x^jv^k + x^kv^j - 2/3 \delta^{jk}
x^iv^i) dm
\,,
\end{equation}
and numerically taking the second time derivative during a
post-processing step.  In \fref{fig:gwtrains} we show the plus
polarization of the GW as seen from an observer on the equator
($h_+^{\mathrm{eq}}$; this is the only non-zero component of the
gravitational waves from axisymmetric simulations) for \mesa{},
\mesapert{}, \mesatwod{}, and \mesatwodpert{} along with the GWs from our
low resolution simulations, \mesalr{}, \mesalrpert{} and our simulation
with velocity dependence in the neutrino transport, \mesavlr{}.

The first observation is that due to the symmetries imposed in 2D,
gravitational wave signal is between 10 and 20 times the strength of
the 3D signal. Note the scale difference, of a factor of 10, for the
top two panels of \fref{fig:gwtrains} compared to the bottom three
panels.  Gravitational waves are generated from large coherent matter
motions in the turbulent gain region and near the PNS surface.  In 2D,
the axisymmetric assumption leads to an artificial enhancement of this
coherent motion. This is in addition to the consequences of the
reverse turbulent cascade in 2D which also generates large scale
structures not seen in 3D simulations.  This observation is in
agreement with other analyses \citep{muller:1997, andresen:2017}
although other comparisons between the gravitational wave signature in
2D and 3D find much less of a suppression in 3D \citep{yakunin:2017}.

\begin{figure*}[tb]
  \plottwo{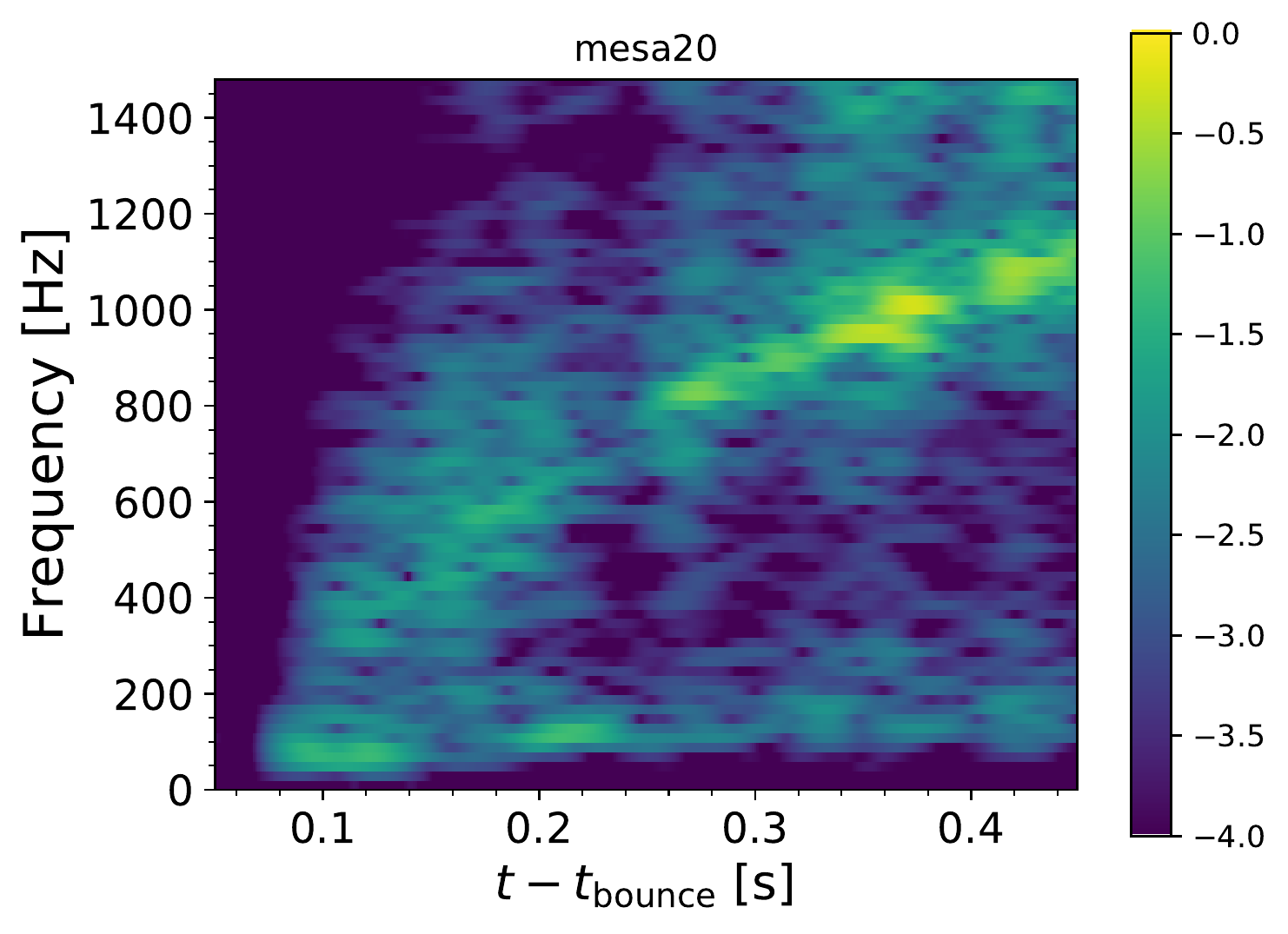}{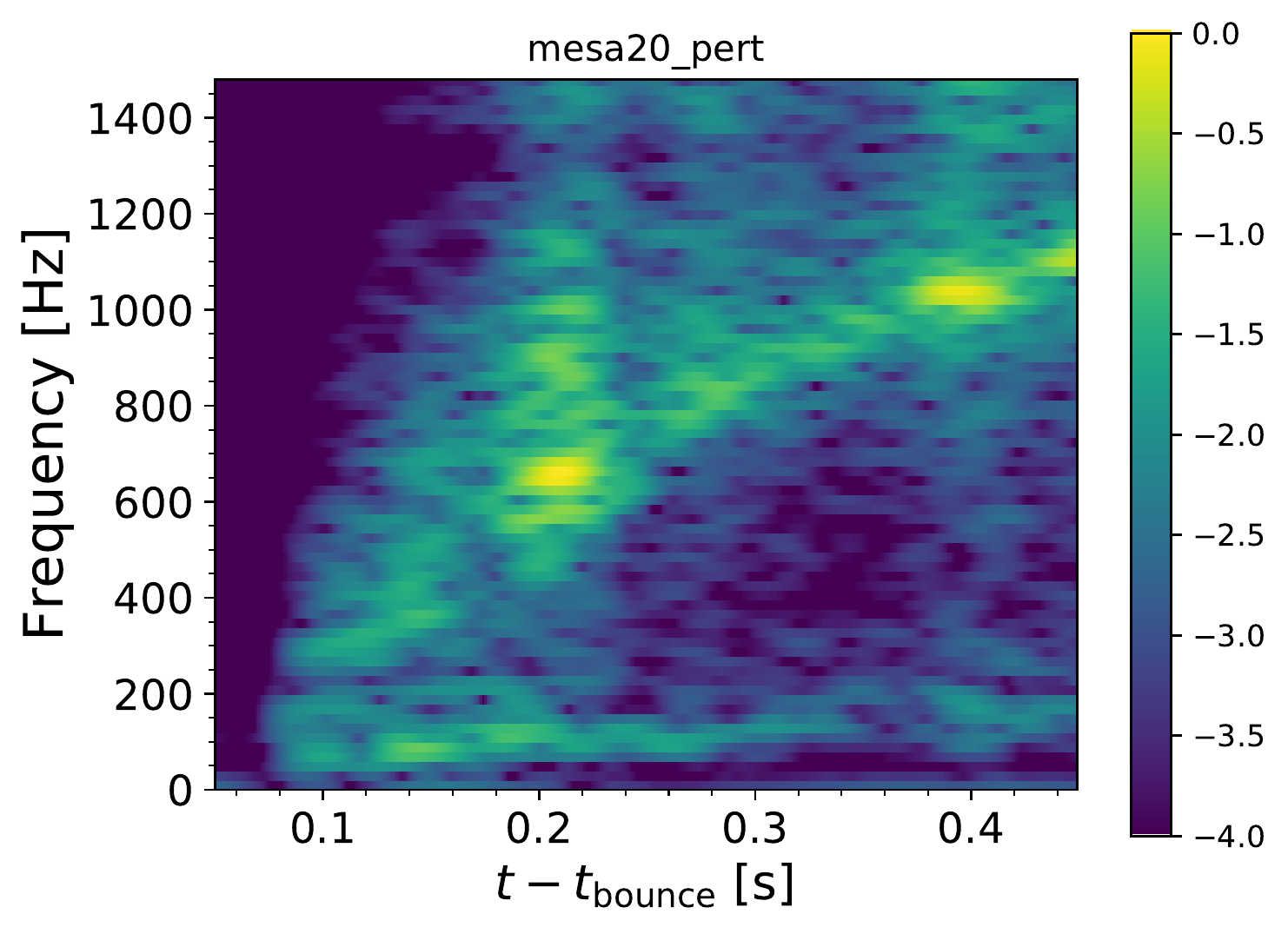}
  \caption{Gravitational wave spectrograms for simulations \mesa{}
    (left) and \mesapert{} (right).  For each time, the gravitational
     wave signal is multiplied by a Hanning window with a width of
    50\,ms.  After the data is Fourier transformed we calculate
    $\log_{10} [|\tilde{h_+}|^2 + |\tilde{h_\times}|^2]$ (where the
    signal are those seen by an observer on the equator) and normalize
    to the same global maximum (across both plots) to allow for a
    direct comparison.}\label{fig:gwspectro}
\end{figure*}

In the third and fourth panel of \fref{fig:gwtrains} we compare the
standard (third panel) and low (fourth panel) resolution simulations
with and without the imposed perturbations.  The most striking impact
of the perturbations is the strong period of GW emission soon after
200\,ms.  At this time the perturbations at the top of the silicon
shell are being accreted through the shock and amplifying the
turbulent motions in the gain region. In addition to boosting the
pressure support behind the shock and, at least temporarily,
increasing the shock radius (see also \sref{sec:perturbations}) these
increased turbulent motions give rise to strong GW emission.  This
effect is also visible in the 2D simulations that include
perturbations (second panel vs. first panel of \fref{fig:gwtrains}).
To locate the source of this enhanced emission we compute spectrograms
of the GW signal.  We show these in \fref{fig:gwspectro} for the
\mesa{} (left panel) simulation and \mesapert{} (right panel)
simulation. The spectrograms generally behave similarly.  There is no
appreciable GW signal until $\sim$100\,ms after bounce when the flow
becomes turbulent.  There are two distinct sources of GWs seen.
First, there is a component that begins at $\sim$300\,Hz are grows
with time.  This component has been associated with the contracting
(but mass-gaining) PNS \citep{marek:2009,muller:2013,kuroda:2017}. The
frequency we observe is slightly higher (by $\sim$20\% at 450\,ms)
than expected based on the analytic predictions in \cite{muller:2013}
because our gravitational wave predictions arise from Newtonian
hydrodynamics are therefore are not either time dilated or
redshifted. The other visible component is the low frequency
($\sim$50-200\,Hz) GW emission which is arising from the convective
and turbulent region behind the shock. The main difference in the GW
spectrograms between the \mesa{} and the \mesapert{} simulations is
the excess emission near 200\,ms.  These spectrograms imply that the
primary frequencies of this enhanced emission in \mesapert{} are close
to $\sim$600\,Hz, which at this time is the characteristic frequency
of the PNS. After the perturbations accrete through the shock and
stimulate the growth of turbulence, they then accrete onto the PNS,
excite it, and lead to the production of GWs.  Since we use only
monopole gravity, we have been cautious in interpreting our GW
signals.  To ensure that the presence of this excess emission
associated with the imposed presupernova perturbations is not a result
of the monopole gravity assumption, we have also performed 2D
simulations using $\ell_\mathrm{max}=16$ in our monopole gravity
solver.  All of the gravity moments higher than $\ell=0$ are based on
the Newtonian gravitational potential.  We also see the excess
gravitational wave emission due to progenitor perturbations in these
simulations.

In addition to the bursts of GWs around 200\,ms due to the imposed
progenitor perturbations, we also see bursts of GW emission associated
with the collapse of spiral SASI waves. In particular, the increased
emission at $\sim$360\,ms in \mesa{} and at $\sim$400\,ms in
\mesapert{}.  While not shown, we see a similar feature in \mesalr{}
at $\sim$370\,ms.  The emission is concentrated at the peak PNS
frequency, but does contribute power to other frequencies as well. In
terms of the dynamics, both the accretion of progenitor perturbations
and the collapsing spiral SASI waves are very similar.  They both
excite turbulence in the gain region marked by an increase in the
lateral kinetic energy (see \fref{fig:mesa20vspert}).

Finally, we compare \mesalr{} and \mesavlr{}.  During the first
$\sim$300\,ms we see broad agreement in the GW signal between these
two simulations.  Starting at $\sim$350\,ms, coincident with the
increased SASI activity, the power in the GW signal of \mesalr{} grows
while \mesavlr{} stays roughly constant.  This is in agreement with our
previous observations which show reduced SASI motions in the \mesavlr{}
simulation due to the larger shock radius and slower recession of the
PNS core (see \fref{fig:vdep}). 

\section{Discussions and Conclusions}
\label{sec:conclusions}

In this study, we have presented the first suite of 3D simulations
using the FLASH hydrodynamics framework with energy-dependent,
multi-dimensional, M1 neutrino transport. We perform eight 3D
simulations exploring the impact of imposed progenitor perturbations,
resolution, octant symmetries, dimensionality, and velocity dependence
in the transport equations. For these simulations we use the same
progenitor with a ZAMS mass of 20$M_\odot$ produced using the MESA
stellar evolution package. We find no explosions in any of our 3D
simulations.

We have presented a thorough analysis of each of the above mentioned
variations. The impact of imposed progenitor perturbations is clear.
We find that the velocity perturbations we place in the silicon shell,
particularly those located near the top of the shell, are very
efficient at seeding turbulence in the post-shock flow.  We find an
enhancement of the lateral kinetic energy when these perturbations
accrete through the shock, a corresponding increase in the neutrino
heating, shock radius, and a quantitive measure of how close the
simulation is to explosion.  Ultimately we aspire to use progenitors
evolved in full 3D (for as long as possible before collapse) in order
to get a more physical set of initial conditions.  We expect
(following \cite{couch:2015a,muller:2017}, and as we see here) that
these precollapse asphericity are crucial in helping an explosion
develop.  Lower resolution, which leads to larger numerical seeds,
slightly raises the level of turbulence as well and gives a slightly
larger shock radius, and a higher measure of the closeness to
explosion, however this effect is smaller than that observed from the
imposed perturbations.  Overall the low resolution simulations behave
qualitatively similar to the higher resolution simulations.  Imposing
octant symmetry places restrictions on the development of global
instabilities, particular the SASI.  It is these lowest order modes
which can play an important role in helping to support the shock at
larger radii which can increase the neutrino heating. These
simulations, which lack these low order modes are quantitatively
further from explosion. We have explored, though a limited number of
simulations, the impact of improved neutrino transport methods.  The
largest difference of these simulations, compared to the simulations
performed without these improvements, is the presence of stronger
protoneutron star convection.  The effect is a larger shock radius (by
30\% at 300\,ms), increased lateral kinetic energy, and an increased
heavy-lepton neutrino luminosity. We discuss additional differences
below.

We also explore in detail the SASI.  We observe this phenomenon in all
of our simulations without velocity dependence.  We mainly observe
spiral SASI modes, although there are non-zero, non-spiral components
as well. Coincident with the SASI modes are correlated variations in
the neutrino signals, showing at most a $\sim$5\% variation on top of
the baseline signal. In many cases the spiral SASI modes build up,
become non-linear, and collapse.  This triggers a burst of turbulence
that increases the neutrino heating and helps support and push the
supernova shock to larger radii. None of these events leads to a
successful explosion, however, each of these events temporarily brings
the core-collapse event quantitatively closer to explosion.  We
suggest that such SASI-driven explosions may be one way the supernova
shock can be reenergized.  In such an event, the spiral SASI wave will
have imparted a non-zero angular momentum to the core, even for
progenitors with little or no initial rotation.  We find that the
typical final neutron star spin period for the spiral SASI waves in
our simulation is $\sim$1\,s. This relatively slow period is in part
due to the small shock radii at the times when we see strong SASI
activity. This limits the amount of angular momentum that can be built
up in the spiral SASI wave. Bearing in mind that we have only
performed one full 3D simulation with velocity dependence (and it was
performed at our lower resolution), we suggest that a consequence of
the increased PNS and shock radii in that simulation is a lack of the
presence of the SASI. This will take further exploration with more
full 3D models with the full neutrino transport and our standard
resolution.

While the one full 3D simulation with velocity dependence does not
exhibit SASI motions, it does show conclusive evidence for the LESA
phenomena that was first reported in \cite{Tamborra:2014}. The work
presented here is the first reported independent confirmation of the
LESA and the first to demonstrate that LESA is not an artifact of the
ray-by-ray method.  In this simulation we find a substantial dipole
component in the total electron lepton number emission which grows
starting at $\sim$200\,ms after bounce and reaches a maximum at
$\sim$300-400\,ms after bounce.  This dipole is stable, yet migratory,
in direction.  It shows excellent alignment with a dipole of the
electron fraction in the convective PNS core.  We confirm many of the
observations seen in \cite{Tamborra:2014} include hemispheric excess
of neutrino heating and a larger average shock radius in the direction
anti-aligned with the lepton-number dipole direction, and a larger
mass accretion rate onto the PNS in the hemisphere aligned with the
lepton-number dipole direction.  It still remains unclear the
mechanism from which this instability is initially generated and
sustained. We conclude that the reason for seeing this only in the one
simulation performed with velocity dependence is not the explicit
presence of this improvement in the neutrino transport, but rather the
impact of the improvement on the strength of PNS convection in the
core.  A stronger PNS convection appears to be conducive to the
appearance of the LESA.

We also investigate the GW signal from our simulations.  We see a
large impact of the imposed progenitor perturbations on the GW
signal. At the time when we see the largest impact on the dynamics,
when the outermost layers of the silicon shell accrete through the
shock, we see a large enhancement of the GW signal, with an amplitude
$\sim$4-5 times greater than the signal from the unperturbed
simulation.  Most of the excess power is at frequencies that are
associated with oscillations of the PNS.  When the aforementioned
spiral SASI waves undergo collapse, we also see bursts of GWs.  These
bursts are coincident with increased turbulent activity in the gain
region.

In summary, our suite of simulations reveals a plethora of 3D dynamics
that altogether play a role in core-collapse supernovae.  We show,
through several mechanisms, that increased turbulent activity, however
it arises, can play a crucial role in the shock dynamics by providing
additional pressure support, facility increased neutron heating, and
quantitatively bring the system closer to the point of explosion.
While we see no explosions in the work presented here, we are
confident that that simulation framework we have developed and
assessed in this article will be invaluable for our future
explorations of 3D simulations of core-collapse supernovae.

\section*{Acknowledgements}

We thank J\'er\^ome Guilet, Hans-Thomas
Janka, Kuo-Chuan Pan, Luke Roberts, and Irene Tamborra for helpful discussions and Kuo-Chuan Pan for assistance with the volume renderings in \fref{fig:vrs}.  Partial support for this
work (E.O.) was provided by NASA through Hubble Fellowship grant
\#51344.001-A awarded by the Space Telescope Science Institute, which
is operated by the Association of Universities for Research in
Astronomy, Inc., for NASA, under contract NAS 5-26555. 
S.M.C. is supported by the U.S. Department of Energy, Office of Science, Office of Nuclear Physics,
under Award Numbers DE-SC0015904 and DE-SC0017955 and the Chandra X-ray Observatory under grant TM7-18005X.
The software used in this
work was in part developed by the DOE NNSA-ASC OASCR Flash Center at
the University of Chicago.  This research used computational resources
at ALCF at ANL, which is supported by the Office of Science of the US
Department of Energy under Contract No. DE-AC02-06CH11357. We
particularly thank
Adrian Pope at ALCF for assistance.  Additional
resources from TACC
under NSF XSEDE allocation TG-PHY100033, and on the Zwicky cluster at
Caltech, which is supported by the Sherman Fairchild Foundation and by
NSF award PHY-0960291 have been used.

\software{FLASH \citep{fryxell:2000, dubey:2009}, GR1D
  \citep{oconnor:2010,oconnor:2015}, Matplotlib \citep{Hunter:2007},
  yt \citep{Turk:2010}, HEALPix/healpy \citep{HEALPix:2005}, MESA
  \citep{paxton:2011, paxton:2013, paxton:2015, paxton:2018}, VisIt
  \citep{HPV:VisIt}}

\end{document}